\documentclass[usenatbib]{aastex}

\received{May 24, 2018}
\revised{April 8, 2019}
\accepted{April 10, 2019}

\usepackage{mathptmx}

\usepackage[T1]{fontenc}
\usepackage{ae,aecompl}

\usepackage{float}

\usepackage[alsoload=astro,separate-uncertainty = true,multi-part-units=single]{siunitx}

\usepackage{savesym}
\savesymbol{splitbox}
\usepackage{adjustbox}
\restoresymbol{TXF}{splitbox}
\savesymbol{captionbox}
\usepackage{support_caption}
\usepackage{caption}
\usepackage{subcaption}
\restoresymbol{TXF}{captionbox}

\usepackage{enumitem,booktabs,cfr-lm}
\usepackage[referable]{threeparttablex}
\renewlist{tablenotes}{enumerate}{1}
\makeatletter
\setlist[tablenotes]{label=\tnote{\alph*},ref=\alph*,itemsep=\z@,topsep=\z@skip,partopsep=\z@skip,parsep=\z@,itemindent=\z@,labelindent=\tabcolsep,labelsep=.2em,leftmargin=*,align=left,before={\footnotesize}}
\makeatother

\usepackage{cleveref}
\usepackage{textcomp}

\usepackage{graphicx}	
\usepackage{amsmath}	
\usepackage{amssymb}	

\shorttitle{Multiwavelength S26}
\shortauthors{Gross et al.}

\begin{document}

\title{Multiwavelength Study of the X-ray Bright Supernova Remnant N300-S26 in NGC 300}


\author{Jacob Gross\altaffilmark{1},
Benjamin F. Williams\altaffilmark{1},
Thomas G. Pannuti\altaffilmark{2},
Breanna Binder\altaffilmark{3},
Kristen Garofali\altaffilmark{4},
Zachary G. Hanvey\altaffilmark{2}
}
\altaffiltext{1}{Department of Astronomy, Box 351580, University of Washington, Seattle, WA 98195}
\altaffiltext{2}{Space Science Center, Department of Earth and Space Sciences, 235 Martindale Drive, Morehead State University, Morehead, KY 40351}
\altaffiltext{3}{Department of Physics and Astronomy, California State Polytechnic University, 3801 West Temple Ave, Pomona, CA 91768}
\altaffiltext{4}{Department of Physics, 825 West Dickson Street, University of Arkansas, Fayetteville, AR 72701}






\label{firstpage}

\begin{abstract}

We present a multiwavelength examination of the supernova remnant (SNR) S$26$ in the nearby galaxy NGC $300$ using data from \textit{Chandra X-ray Observatory}, \textit{XMM-Newton X-ray Observatory}, \textit{Hubble Space Telescope} (HST), the \textit{Very Large Array}, and the \textit{Australia Telescope Compact Array}. We simultaneously fit all of the available X-ray data with a thermal plasma model and find a temperature of $0.77 \pm 0.13$ keV with a hydrogen column density of ($9.7^{+6.4}_{-4.8}$)$\times 10^{20}$ cm$^{-2}$. HST imaging allows us to measure a semimajor axis of $0.78 \pm 0.10$ arcsec ($7.5 \pm 1.0\ \parsec$) and a semiminor axis of $0.69^{+0.14}_{-0.12}$ arcsec ($6.7^{+1.2}_{-1.4}\ \parsec$). This precise size helps to constrain the age and velocity of the shock to be ($3.3^{+0.7}_{-0.6}$)$\times 10^{3}$ yr and $411^{+275}_{-122}$ km s$^{-1}$. We also fit photometry of the surrounding stars to infer the age and mass of the progenitor star to be \SI{8 \pm 1}{\mega yr} and $25^{+1}_{-5}$ M$_{\odot}$. Based on measured radio properties of the source and assuming equipartition, the estimated radio luminosity of $\sim 1.7\times 10^{34}$ erg s$^{-1}$ over the $10^{8}$ to $10^{11}$ Hz frequency range results in a minimum magnetic field associated with this SNR of $0.067$ mG and the minimum energy needed to power the observed synchrotron emission of $1.5\times 10^{49}$ erg. The size and temperature of N$300$-S$26$ appear to be similar to the Galactic SNR G$311.5$--$0.3$ except that G$311.5$--$0.3$ has a significantly lower X-ray luminosity, is older, and has a slower shock velocity.

\end{abstract}




\section{Introduction}
\label{intro}

The energy output of supernova remnants (SNRs) shock and excite the interstellar medium (ISM), which makes them visible across the electromagnetic spectrum out to distances of megaparsecs. While the Milky Way SNR population is closest to us, many of its SNRs are difficult to observe due to interstellar absorption along Galactic lines of sight and uncertainties in distance measurements. Analyses of SNRs outside our Galaxy provide key comparisons to the Galactic sample that probe differences in the ISM and how it influences SNR's multiwavelength morphological luminosity evolution. Since the nearby galaxy population contains a wide range of ISM densities and metallicities, detailed multiwavelength measurements of the SNRs allow us to shed light on the effects of the environment on SNR properties.

There have been many SNR surveys of nearby galaxies including the Large Magellenic Cloud (LMC) \citep{Desai2010,Seok2013,Bozzetto2017}, M$31$ \citep{Kong2003,Jennings2014,Lee2014M31}, M$33$ \citep{BlairLong2010,Jennings2014,Lee2014M33}, and M$83$ \citep{Blair2012,Long2014,Winkler2017}. One such galaxy is NGC $300$, a face-on spiral ScD galaxy at \SI{2}{\mega\parsec} \citep{Dalcanton2009}, which is the brightest of the five main spiral galaxies in the Sculptor Group \citep{Rodriguez2016}. The $46^{\circ}$ inclination of NGC $300$ \citep{Freedman1992} as well as the Galactic latitude of $-79.4$ degrees (which places NGC $300$ toward the southern Galactic pole; \cite{Dalcanton2009}) allows it to be observed easily due to reduced absorption effects from gas in the host galaxy as well as our Galaxy. There are many different surveys of NGC $300$'s SNR population including, but not limited to, optical \citep{DOdorico1980, BlairLong1997, Millar2011}, radio \citep{Pannuti2000}, and X-ray \citep{Payne2004,Carpano2005}. All of this data allows for a detailed measurement of the physical properties of SNRs in NGC $300$.

Multiwavelength observations of SNRs provide details about the energetics and evolution associated with these sources. The X-ray, optical, and radio emission from SNRs probe the magnetic fields created by the source, the energy needed to drive synchrotron radiation, the temperature and column density of the surrounding gas, the physical size and velocity of the associated shock waves, and many other properties. These SNR properties should, in principle, be related to both the physical parameters of the progenitor star and those of the surrounding ISM.

We have observed a bright extragalactic SNR in NGC $300$ denoted as N$300$-S$26$ (\cite{DOdorico1980,BlairLong1997}; hereafter referred to as S$26$). Previous observations of S$26$ have included various radius measurements taken from multiple optical ground-based telescopes ($1.7$ arcsec which corresponds to $16.5\ \parsec$ at the assumed distance to NGC $300$; \cite{DOdorico1980,BlairLong1997}), H$\alpha$ surface brightness \citep{BlairLong1997}, radio flux density \citep{Pannuti2000,Payne2004}, as well as a temperature from the \textit{ROSAT} telescope \citep{Read1997}. 

In this paper, we have observed S$26$ in the optical using \textit{Hubble Space Telescopes} (\textit{HST}), in the X-ray by \textit{Chandra} and \textit{XMM-Newton}, and in the radio using the \textit{Very Large Array} (\textit{VLA}) and \textit{Australia Telescope Compact Array} (\textit{ATCA}) from \cite{Pannuti2000} and \cite{Payne2004}. We then compare S$26$ to other SNRs in the Galaxy, finding that S$26$ is most similar to G$311.5$--$0.3$, which has a temperature of $0.68^{+0.20}_{-0.24}$ keV and a radius of \SI{9}{\parsec}. In \cref{sec:obs_red}, we discuss the data sets used in our study and the methods utilized to extract information from the raw data. In \cref{sec:results}, we share our results from our data reductions. In \cref{sec:discussion}, we discuss the physical ramifications of our measurements. In \cref{sec:conclusion}, we summarize our results.

Values derived in the radio sections are at the $90$\% confidence range. The values derived in the X-ray sections for the {\tt tbabs*(pshock)} normalization tied model are at the $90$\% confidence range while the other models are at the $1 \sigma$ limit. 

\section{Observations and Data Reduction}
\label{sec:obs_red}

Our multiwavelength program makes use of data from the \textit{HST} (optical), the \textit{Chandra X-ray Observatory} (\textit{Chandra}; X-ray), the \textit{XMM-Newton X-ray Observatory} (\textit{XMM-Newton}; X-ray), the \textit{VLA} (radio), and the \textit{ATCA} (radio) (radio data from \cite{Pannuti2000} and \cite{Payne2004}). We now discuss each of these in detail below.

\subsection{Optical Data}
\label{sec:optical}

We detected the optical shell of S$26$ with the Advanced Camera for Surveys on board the \textit{HST} at $0$:$55$:$15.447$, $-37$:$44$:$39.10$ (J$2000$). The data were obtained with the WFC detector using F$814$W and F$606$W filters on $2015$ January $19$. The F$814$W filter had a $966$ s exposure and the F$606$W filter had a $850$ s exposure. While the broadband SNR fluxes would not be useful for science, as they contain multiple emission lines and have a dense background of stars, the high-resolution imaging allowed us to measure a precise size for the optical shell.  In addition, the field of resolved stars allowed us to perform crowded stellar field photometry on the stellar population local to the SNR, which provides constraints on the mass of the star that produced the SNR.  We detail the analysis techniques we employed for each of these applications below.

\subsubsection{Optical SNR Size Measurement}

\begin{figure*}[!htb]
    \adjustbox{valign=b}{%
    \begin{minipage}[t]{.5\linewidth}
    	\begin{subfigure}{\linewidth}
        \centering
        \includegraphics[width = 3.0in]{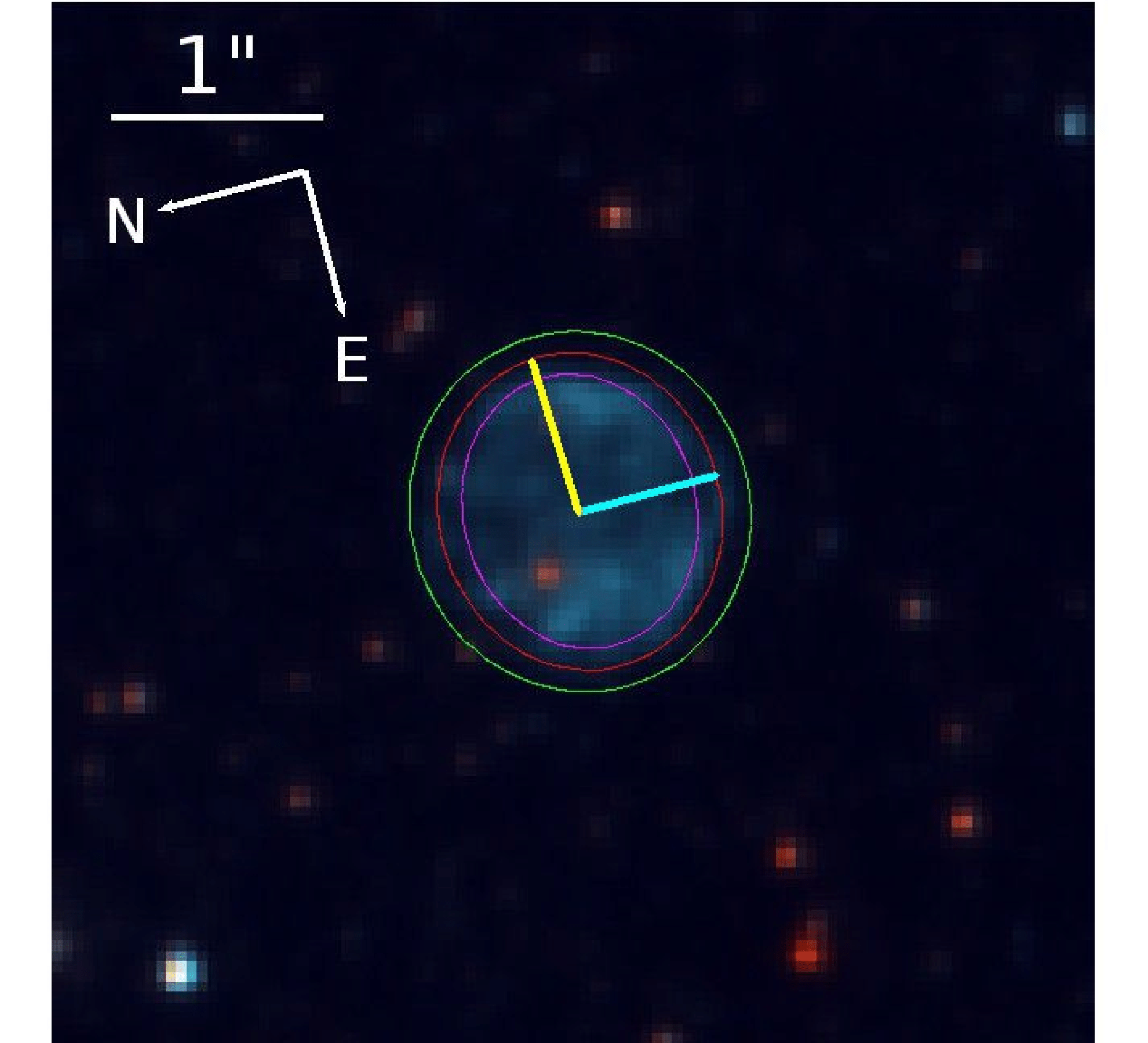}
        \end{subfigure}\par\bigskip
    	\begin{subfigure}{\linewidth}
        \centering
        \includegraphics[width = 2.8in]{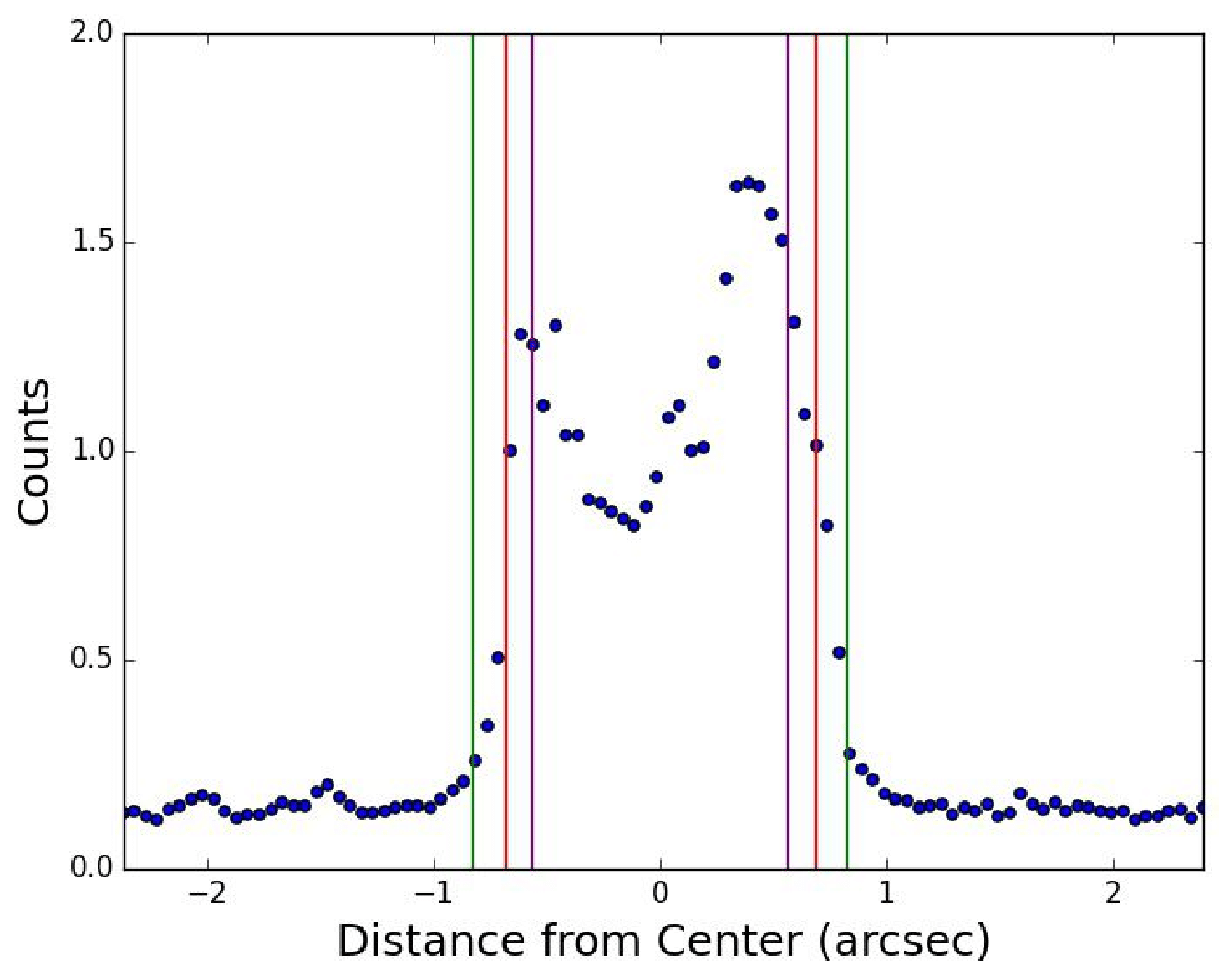}
        \end{subfigure}\par\bigskip
        \begin{subfigure}{\linewidth}
        \centering
        \includegraphics[width = 2.8in]{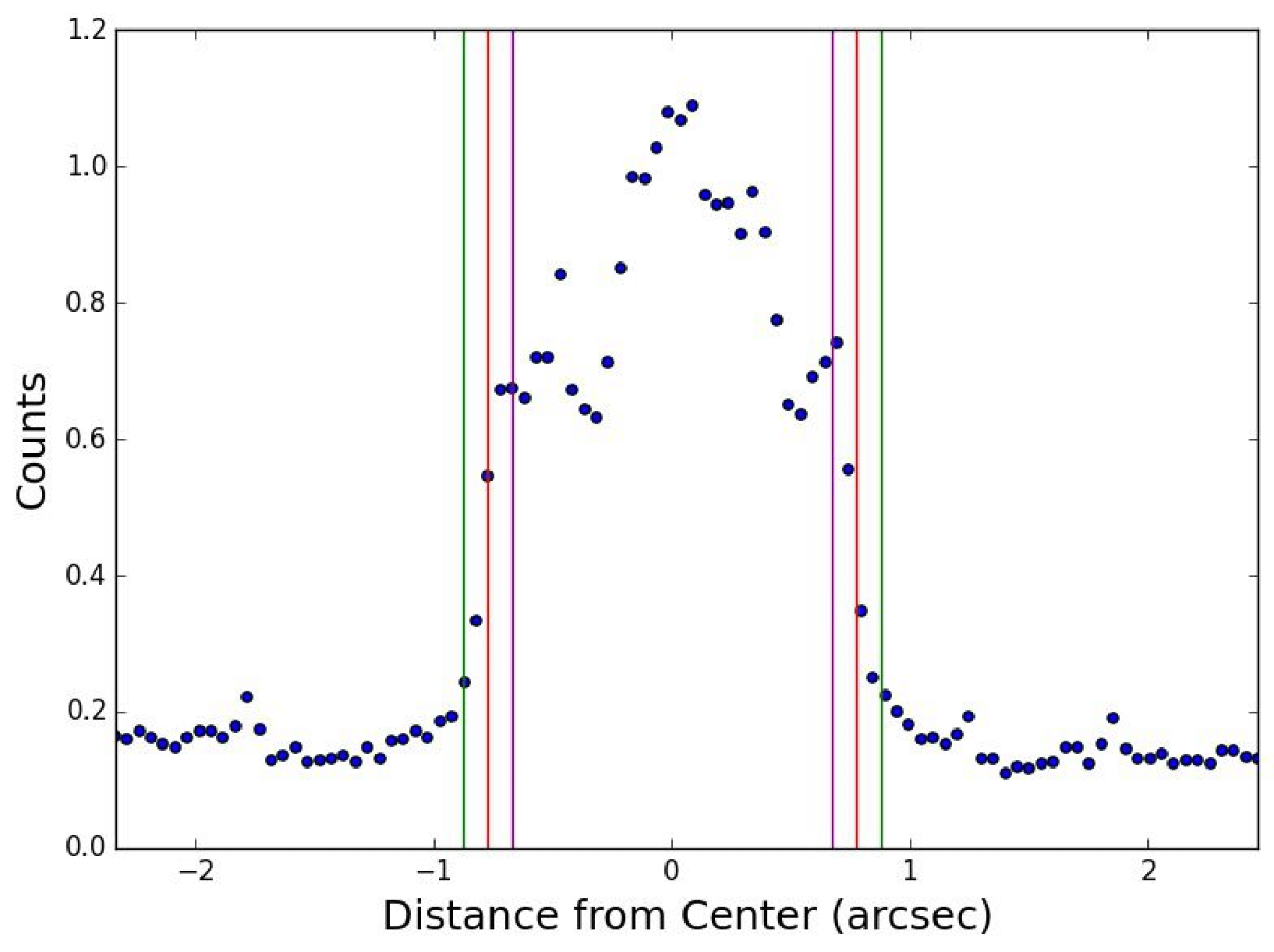}
        \end{subfigure}
    \end{minipage}}%
    \adjustbox{valign=b}{%
    \begin{minipage}[t]{.5\linewidth}
    \caption{\small \\\hspace{\textwidth}\textbf{Top:} Optical image of S$26$ using F$606$W and F$814$W filters as well as an estimated blue channel using $2\ \times$ F$606$W - F$814$W. The semiminor axis is depicted with the blue line while the semimajor axis is depicted with the yellow line.\\\hspace{\textwidth}\textbf{Middle:} Semiminor measurement using Surface Brightness Function aligned with semiminor axis taken from the optical image.\\\hspace{\textwidth}\textbf{Bottom:} Semimajor measurement using Surface Brightness Function aligned with semimajor axis taken from the optical image.\\\hspace{\textwidth}\\\hspace{\textwidth}Green ellipse and vertical lines correspond to upper limit to fit. Red ellipse and vertical lines correspond to best-fit. Purple ellipse and vertical lines correspond to the lower limit to fit.}
    \label{fig:opt_size}
    \end{minipage}}%
\end{figure*}

We used Deep Space $9$ (DS$9$) version $7.2.1$\footnote{http://ds9.si.edu/site/Home.html} to create an RGB-rendered image of the source using F$606$W for the green channel, F$814$W for the red channel, and an estimated blue channel of $2\ \times$ F$606$W $-$ F$814$W (Figure~\ref{fig:opt_size}). We then used ellipses to measure the size of the SNR to a significantly higher precision than previous studies due to the \textit{HST}'s superior spatial resolution (see \cref{sec:results_size} for details).

In order to corroborate the size measurement using ellipses, we used the DS$9$ tool {\tt Projections} to create surface brightness functions along the semimajor and semiminor axes.

\subsubsection{Resolved Stellar Photometry}

We also measured resolved stellar photometry from the \textit{HST} images in order to produce color-magnitude diagrams (CMDs) of the stellar populations surrounding the SNR using the VEGAMAG system. The photometry was performed using the automated point spread function (PSF) fitting pipeline from the Panchromatic Hubble Andromeda Treasury.  The full details of how the pipeline works are given in \cite{Williams2014}, but briefly, the calibrated flat-fielded and CTE corrected ({\tt flc}) \textit{HST} images are masked and analyzed using a combination of the PyRAF routine {\tt astrodrizzle} and the photometry package DOLPHOT, an updated version of HSTPHOT \citep{Dolphin2000}.  The analysis is performed on the full set of images simultaneously, where the locations of stars are found using the statistics of the full stack of images, and the photometry is performed through forcing PSF fitting at all of the star locations on all of the individual exposures.  The resulting measurements are combined and culled based on signal to noise and measurement quality.  We then perform a series of artificial star tests, whereby a star of known color and magnitude is inserted into the images and the photometry routine is rerun to assess whether the star was recovered, and how the output magnitude compared to the input.  This exercise is repeated $10^{5}$ times to build up statistics on completeness, photometric bias, and photometric error, as a function of color and magnitude.  For example, the artificial stars showed that our completeness falls below $50$\% at F$606$W=$27.9$ and F$814$W=$27.1$, and the uncertainties at those magnitudes is $27.9^{+0.4}_{-0.2}$ and $27.1^{+0.3}_{-0.2}$ for F$606$W and F$814$W, respectively.  The asymmetric uncertainties are due to the bias of faint stars being measured brighter than their true flux due to crowding effects. 

\subsection{X-Ray Data}
\label{sec:xray}

The X-ray data used were three \textit{Chandra} and six \textit{XMM-Newton} observations. The information about each observation can be seen in \Cref{tab:xray_obs}. The SNR is near two other sources, as shown in Figure~\ref{fig:xray_extraction}; however, it is separated enough from these ($30.1$ and $50.1$ arcsec for the two objects) that we were able to mask them out as shown in the figure. We discuss the details of these observations below.

\subsubsection{Chandra Data}
\label{sec:xray_chandra}

\begin{figure*}[!htb]
	\centering
    \includegraphics[height = 2.9in]{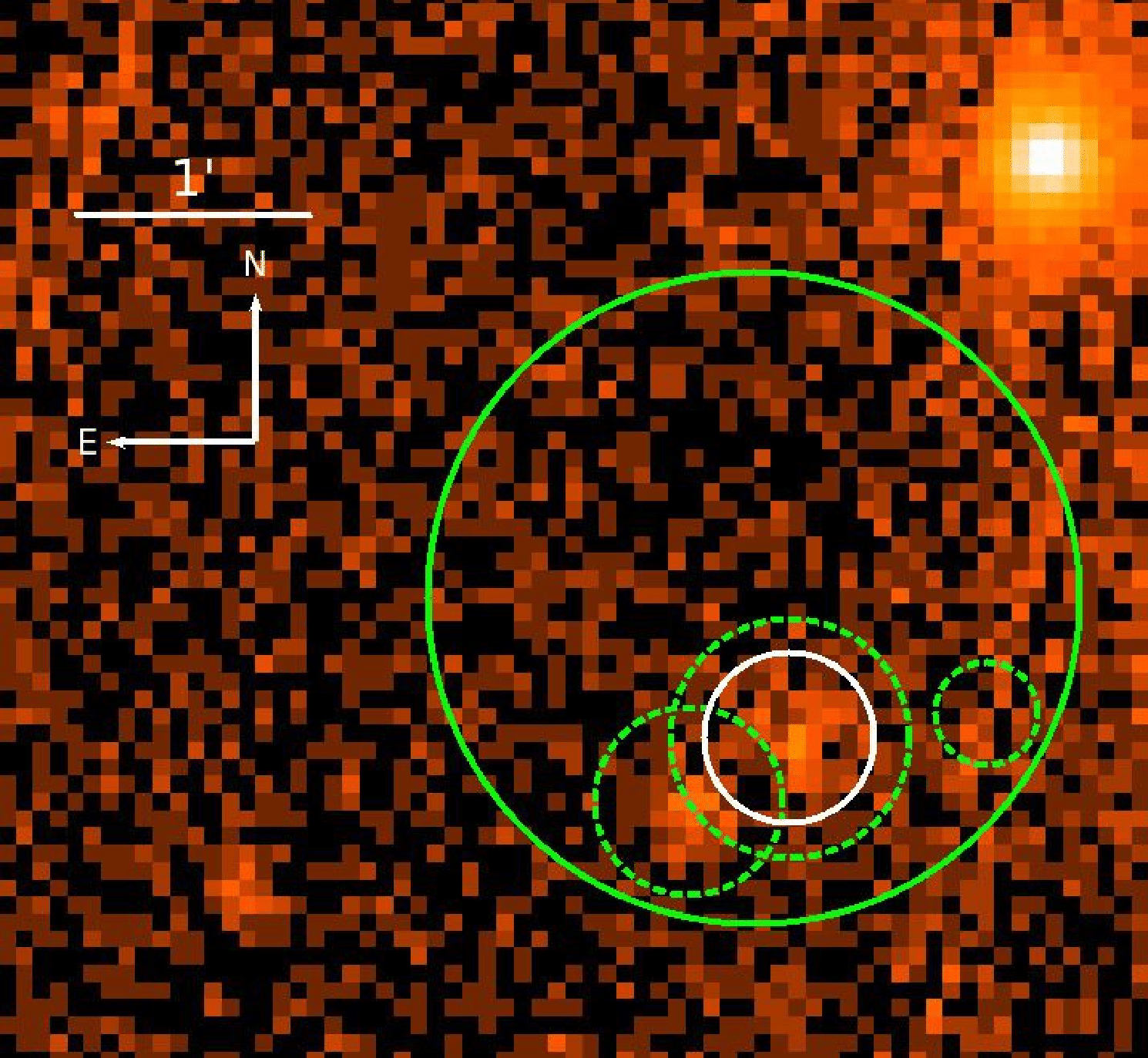}
    \includegraphics[height = 2.9in]{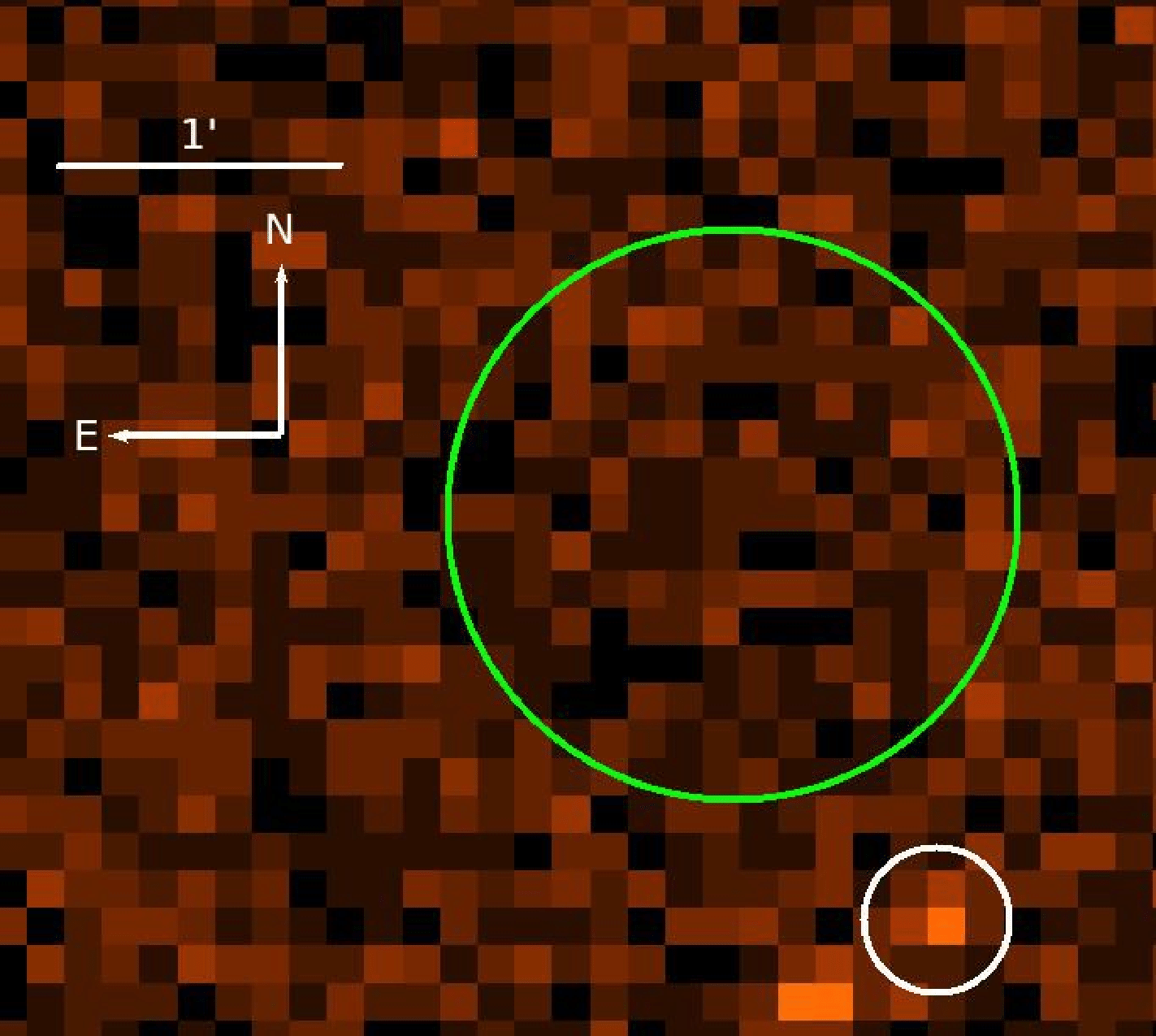}
    \caption{\small \\\hspace{\textwidth}\textbf{Left:} X-ray image of the PN detector for the \textit{XMM-Newton} observation $0112800101$ over the $0.2-3.0$ keV energy range. The large green circle has a radius of $82.7$ arcsec, the green dotted circles have a radius of $23.9$, $30.5$, and $13.1$ arcsec from left to right, and the white circle has a radius of $21.8$ arcsec.\\\hspace{\textwidth}\textbf{Right:} X-ray image of the \textit{Chandra} observation $12238$ over the $0.2-3.0$ keV range. The green circle has a radius of $59.8$ arcsec while the white circle has a radius of $15.4$ arcsec.\\\\\hspace{\textwidth}Source extraction region is depicted as the white circle. Background extraction regions are depicted as the green circles with the dashed circles corresponding to regions that were masked from the background extraction region due to the sources within. The larger circle for the background extraction region was shifted for each observation and each detector for the \textit{XMM-Newton} data in order to minimize the amount of the chip gap in the background region. For both of these images, we used a bin-size of $16$ detector pixel, a min-max log scale, and a Gaussian smoothing of radius $3$ in DS$9$.}
    \label{fig:xray_extraction}
\end{figure*}

We extracted the spectroscopic data using Ciao v$.4.6.7$\footnote{http://cxc.harvard.edu/ciao/} and CALDB v$4.1$\footnote{http://cxc.harvard.edu/caldb/}. Since the source was unresolved in X-rays, we adopted the point-source extraction method for the spectrum from the {\tt specextract} command\footnote{http://cxc.harvard.edu/ciao/threads/pointlike/}. We have a total of three observations using the ACIS detector totaling an exposure of $191$ ks (see \Cref{tab:xray_obs}). For this, we selected regions centered on S$26$, which enclosed the entire source within a circle of radius $15.4$ arcsec. The background region was centered around a nearby empty patch of sky that had no sources within a circle of radius $59.8$ arcsec such that the area for the background region was $\sim 15$ times larger than the area for the source region. The source region was limited to only $15.4$ arcsec because we wanted to maximize the extraction region yet there were other nearby sources that had to be avoided and not just masked out (see \Cref{fig:xray_extraction}).

\begin{table*}[!htb]
	\centering
    \resizebox{\textwidth}{!}{
    \begin{tabular}{ccccccc}
    	\hline
        \hline
                   & Detector                    & Date         & Off-axis Angle & Filtered           & Counts           & Background Fit \\
        Obs ID     & (With Filter)               & (yyyy mm dd) & (arcmin)       & Exposure Time (ks) & (0.3 to 2.0 keV) & Cash Statistic/dof \\
        \hline
        12238      & ACIS-I VFAINT               & 2010 Sep 24  & 7.180          & 63.8               &  50             & 756/661\\
        16028      & ACIS-I FAINT                & 2014 May 16  & 5.950          & 65.1               &  49             & 168/148\\
        16029      & ACIS-I FAINT                & 2014 Nov 17  & 5.286          & 62.1               &  49             & 188/148\\
        \hline
                   & EPIC-MOS$1$ Medium          &              &                & 42.9               &  79              & 437/537\\
        0112800101 & EPIC-MOS$2$ Medium          & 2001 Jan 2   & 5.448          & 42.3               &  67              & 436/536\\
                   & EPIC-PN Medium              &              &                & 35.6               &  175             & 525/540\\
        \hline
                   & EPIC-MOS$1$ Medium          &              &                & 27.5               &  37              & 355/537\\
        0112800201 & EPIC-MOS$2$ Medium          & 2000 Dec 27  & 5.448          & 29.9               &  65              & 355/536\\
                   & EPIC-PN Medium              &              &                & 24.5               &  175             & 515/540\\
        \hline
                   & EPIC-MOS$1$ Medium          &              &                & 35.0               &  57               & 437/537\\
        0305860301 & EPIC-MOS$2$ Medium          & 2005 Nov 25  & 5.812          & 35.0               &  67               & 435/536\\
                   & EPIC-PN Medium              &              &                & 33.0               &  200              & 536/540\\
        \hline
                   & EPIC-MOS$1$ Medium          &              &                & 24.0               &  37               & 363/537\\
        0305860401 & EPIC-MOS$2$ Medium          & 2005 May 22  & 5.812          & 23.0               &  42               & 363/536\\
                   & EPIC-PN Medium              &              &                & 22.0               &  128              & 531/540\\
        \hline
                   & EPIC-MOS$1$ Thin$1$         &              &                & 91.0               &  193             & 318/346\\
        0791010101 & EPIC-MOS$2$ Medium          & 2016 Dec 17  & 1.781          & 98.0              &  188             & 319/345\\
                   & EPIC-PN Medium              &              &                & 91.0               &  741             & 537/495\\
        \hline
                   & EPIC-MOS$1$ Thin$1$         &              &                & 42.0               &  91              & 173/208\\
        0791010301 & EPIC-MOS$2$ Medium          & 2016 Dec 19  & 1.719          & 45.0               &  109             & 177/207\\
                   & EPIC-PN Medium              &              &                & 37.0               &  288             & 362/408\\
        \hline
        13515      & ACS-WFC F$606$W             & 2014 Jun 30  & 1.138          & 2.4                & ...              & ... \\
                   & ACS-WFC F$814$W             &              &                & 2.5                & ...              & ... \\
        \hline
    \end{tabular}
    }
    \caption{\small Observational information of S$26$ in X-Ray and optical energy bands}
    \label{tab:xray_obs}
\end{table*}

Initially, we reprocessed the data from evt$2$ using the Ciao command {\tt chandra\_repro}\footnote{http://cxc.harvard.edu/ciao/ahelp/chandra\_repro.html} using the default settings---observation $12238$ had to be reprocessed with check\_vf\_pha set to true in order to properly reprocess the VFAINT data while the other two observation had check\_vf\_pha set to false. We then used the region files described above in {\tt specextract}. The binning used for both the source and background files was the standard binning practice for {\tt specextract}; namely, the data was grouped in such a way so each bin had a minimum of $15$ counts. 

\subsubsection{XMM-Newton Data}
\label{sec:xray_xmm}

We reduced the \textit{XMM-Newton} data using SAS v$15.0$\footnote{http://xmmssc-www.star.le.ac.uk/SAS/} and using the {\tt xmmselect} command\footnote{https://heasarc.gsfc.nasa.gov/docs/xmm/abc/node9.html}. We have a total of six observations using the MOS$1$, MOS$2$, and PN detectors for each observation that totaled an effective exposure of $262.4$ ks for MOS$1$, $273.2$ ks for MOS$2$, and $243.1$ ks for PN (see \Cref{tab:xray_obs}). 

We created an image of the patch of sky that contained our source while applying filters and flags designed to remove artifacts. For the PN, MOS$1$, and MOS$2$ detectors, we selected the events with PATTERN in the $0-4$ range, set PI to be the preferred pulse height of the event with the range being between $200$ and $3000$ eV, and set $0$xfa$0000$ to $0$ to further clean up the images. We also used the standard practice of setting the \#XMMEA\_EP flag for the PN detector and \#XMMEA\_EM for the MOS$1$ and MOS$2$ detectors.


To filter time intervals with high background counts, we extracted light curves from the three detectors in each of the six observations over the $0.2-3.0$ keV energy range. The command {\tt gtibuild} was used to create Good Time Intervals (GTIs) for each observation, which filtered out data that had any sharp peaks in the light curve. Then, {\tt evselect} was used to filter the data based on the GTIs created with {\tt gtibuild}. The sharp peaks in the light curves were removed in order to obtain the data that corresponded to the unflared time intervals which assists with our spectral fits that were taken.

We then extracted the spectra for our source and background by using the filtered event files via the {\tt xmmselect} command. The {\tt xmmselect} command would then create a PI file and a filtered image for both our source and the background regions. 


The source region used was a circle centered on the SNR with a radius of $21.8$ arcsec that was limited to this size due to nearby sources (see \Cref{fig:xray_extraction}). The background region was chosen near the SNR, avoiding other nearby sources, and any chip gaps. The area of the background region was the same for each observation and each detector and was $\sim 11.4$ times larger than the source region area.

After the source and background spectra were extracted, we created the associated RMF and ARF files using the {\tt rmfgen} and {\tt arfgen} commands. We then ran {\tt grppha} to group the files together similar to the grouping for the \textit{Chandra} data, which used {\tt dmgroup}, but with these bins having a minimum of 1 count to eliminate any empty bins.

\subsubsection{X-Ray Spectral Analysis}
\label{sec:xray_spectral}

After the spectra were extracted, we fit them using XSPEC v$.12.9.0$n\footnote{https://heasarc.gsfc.nasa.gov/xanadu/xspec/}.

Following the technique of \cite{Garofali2017}, we first fit both a sky and instrument background model to our background spectra. This consisted of a pair of absorbed thermal plasma components as the sky background for both the \textit{XMM-Newton} and \textit{Chandra} data.

The instrument background model was different for each of the detectors. For the PN detector, we used a broad Gaussian at $0$ keV and a broken power law, while for the MOS$1$ and MOS$2$ detectors we used a pair of broken power laws. For \textit{Chandra}, we used a combination of power laws and Gaussians. For more information about the background model, see \cite{Garofali2017}.

After we acquired the best-fit background model, we fit the data for the source including the background model components and setting the sky and instrument background parameters to the fitted values. We scaled the normalization of the background by the relative size of the source data region to the background data region. We also included {\tt tbabs} and {\tt pshock} components when fitting the source with a metal abundance of $0.5$ relative to solar abundance (calculated using the metallicity gradient found from \cite{Gazak2015} and using a distance of $3.7$ kpc between S$26$ and the center of NGC $300$)---this metal abundance was accounted for in the {\tt pshock} model component and not the {\tt tbabs} component. The {\tt tbabs} component modeled the interstellar absorption along the line of sight\footnote{https://heasarc.gsfc.nasa.gov/xanadu/xspec/manual/node251.html} assuming a minimum value equal to the Galactic $N_{H}$ (foreground value of $3.0 \times 10^{20}$ cm$^{-2}$ from COLDEN\footnote{http://cxc.harvard.edu/toolkit/colden.jsp}) and any addition to that value being due to the column density from NGC $300$, while the {\tt pshock} component modeled X-ray emission from a constant temperature plane-parallel shock plasma\footnote{https://heasarc.gsfc.nasa.gov/xanadu/xspec/manual/node206.html}.

We fit the six \textit{XMM-Newton} observations and the three \textit{Chandra} observations simultaneously to maximize the amount of counts for the SNR and restricted the energy range to be between $0.3$ and $2.0$ keV---including any higher energy data would just add the background noise because there was no significant detection above $2.0$ keV for S$26$. The only free parameters for the source model were the column density, temperature, and normalizations. The normalization for the source data for each observation was initially free, but the column density and temperature were the same for each observation because we wanted to allow the fit to normalize each observation separately to improve the fit. We also attempted fitting with the normalizations the same for each observation, with the normalizations the same but changed the energy range to be between $0.3$ and $5.0$ keV, and with the normalizations the same but the temperatures allowed to vary from observation to observation and detector to detector. There was also a model that included a {\tt powerlaw} component with the normalizations freed and a model that only consisted of {\tt tbabs} and {\tt powerlaw} components with the normalizations freed. All of these various fits utilized Cash statistics (see \cite{Cash1979}).

\subsection{Radio Data}
\label{sec:radio}

Radio observations of S$26$ from \cite{Pannuti2000} and \cite{Payne2004} using data from \textit{VLA} and \textit{ATCA} have revealed a counterpart to the optically detected SNR. The data was obtained on $1993$ May $22$ for the $6$ cm wavelength data and on $1998$ June $13$ for the $20$ cm wavelength data. The beam size was $\approx 4"$ at $6$ cm (\SI{4885}{\mega\hertz}) and $\approx 6"$ at $20$ cm (\SI{1465}{\mega\hertz}) with an rms sensitivity of \SI{36}{\micro Jy} at $6$ cm and \SI{60}{\micro Jy} at $20$ cm. S$26$ was detected at $20$ cm, but a counterpart was not detected at $6$ cm---\cite{Pannuti2000} and \cite{Payne2004} measured roughly the same value for the flux density of the counterpart at $20$ cm (namely $0.22$ mJy). From these measurements, both papers gave a value for the radio spectral index $\alpha$ (defined such that flux density $S_{\rm{\nu}}$ $\propto$ $\nu^{-\alpha}$) to be $>0.70 \pm 0.05$ (for the purpose of this paper we adopted a lower limit to $\alpha$ of $0.65$ and just calculated the various physical parameters using this limit), which is consistent with synchrotron emission. The \textit{VLA} and \textit{ATCA} observations lacked the angular resolution to resolve clearly any spatial structure in the radio counterpart to S$26$. In \Cref{sec:results_radio}, we discuss how this data was utilized to calculate the minimum energy needed to drive synchrotron radiation and the minimum strength of the magnetic field.

\section{Results}
\label{sec:results}

Using the data extracted over the various wavelengths, we are able to measure a variety of physical parameters. From the optical data taken with \textit{HST}, we can constrain the size of the SNR's shock as well as the mass and age of the progenitor star. From the X-ray data taken with \textit{Chandra} and \textit{XMM-Newton}, we can fit the X-ray spectra with various models to measure the best-fit temperature of the SNR as well as the density of surrounding gas using the normalization factor. From the radio data, we can measure the minimum magnetic field strength as well as the minimum energy needed to drive synchrotron radiation. All of these measurements are described in detail below.

\subsection{Size from HST Data}
\label{sec:results_size}

\begin{figure*}[!htb]
	\centering
	\includegraphics[width = 4.0in]{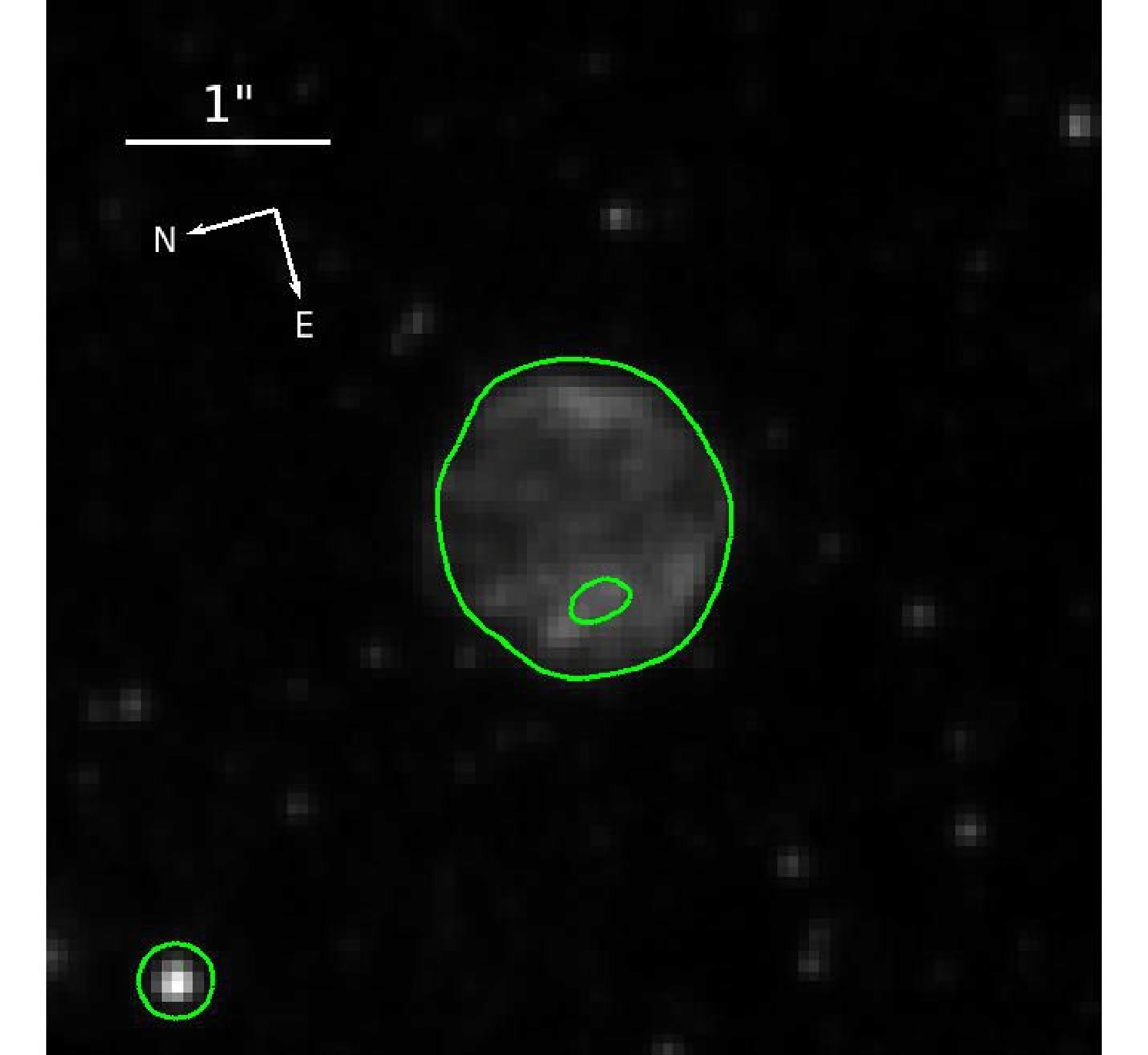}
    \caption{\small Image of the F$606$W \textit{HST} data with the contour used to determine the middle part of the shell edge (i.e. the best-fit value for the size of the SNR).}
    \label{fig:opt_contour}
\end{figure*}

We positioned ellipses by eye to a color image of S$26$ to find the size of the SNR; the color image is shown in \Cref{fig:opt_size}. The ellipse most likely to correspond to the size of S$26$ followed the middle of the shell edge. The lower limit was set to the inner edge of the shell while the upper limit was set to the outer edge of the shell. The edge of the shell was determined by using the DS$9$ tool {\tt Contours} and adjusting the contour level to find the inner, outer, and middle parts of the shell edge for our ellipses that were placed by eye (see \Cref{fig:opt_contour} for the middle part of the shell edge contour). The semimajor axis is $0.78 \pm 0.10$ arcsec and the semiminor axis is $0.69^{+0.14}_{-0.12}$ arcsec, corresponding to $7.5 \pm 1.0\ \parsec$ and $6.7^{+1.2}_{-1.4}\ \parsec$, respectively. We also found that S$26$ has an eccentricity of $0.47$.

We corroborated the ellipse method by also calculating the radii using the DS$9$ tool {\tt Projections} aligned with the semimajor and semiminor axes (see \Cref{fig:opt_size}). This tool can be used to find the surface brightness function for the projected line. The semimajor and semiminor axes' brightness profiles give us a plot of the brightness versus distance that were used to find the semimajor and semiminor axes values---$0.79^{+0.19}_{-0.14}$ arcsec and $0.71^{+0.18}_{-0.24}$ arcsec, respectively. These values were found after performing third-order Taylor series expansions on the outer edges of the surface brightness functions. The values calculated are within the errors from the ellipse measurements.

Previously, the size was estimated to have a diameter of \SI{33}{\parsec} using a distance assumption of \SI{2.1}{\mega\parsec} to NGC $300$ \citep{BlairLong1997}. The \textit{HST} data shows the SNR is a factor of $\sim 2$ times smaller, which is likely due to the improved spatial resolution---\cite{BlairLong1997} had a resolution of $\sim 1"$ while the data from \textit{HST} has a resolution of $\sim 0.1"$.

\subsection{Mass of the Progenitor}
\label{sec:results_mass}

We use the well-established technique of fitting the CMD of the resolved stellar populations within $50$~pc ($5.2"$) of the SNR with stellar evolution models using the fitting package MATCH \citep{Dolphin2000,Dolphin2012,Dolphin2013} to constrain the age of the SNR progenitor \citep[e.g.,][]{Badenes2009,Gogarten2009,Jennings2012,williams2014a,maund2017}. We begin by assuming that the progenitor of S$26$ was a massive star ($>7$~M$_{\odot}$) that underwent core-collapse.  In addition, we assume that nearby young stars were associated with the progenitor star.  With these assumptions, we can use the ages of the nearby stars, as determined from their CMD to measure the most likely age (and inferred mass) of the progenitor.

\begin{figure*}[!htb]
    \centering
	\includegraphics[width=0.8\linewidth]{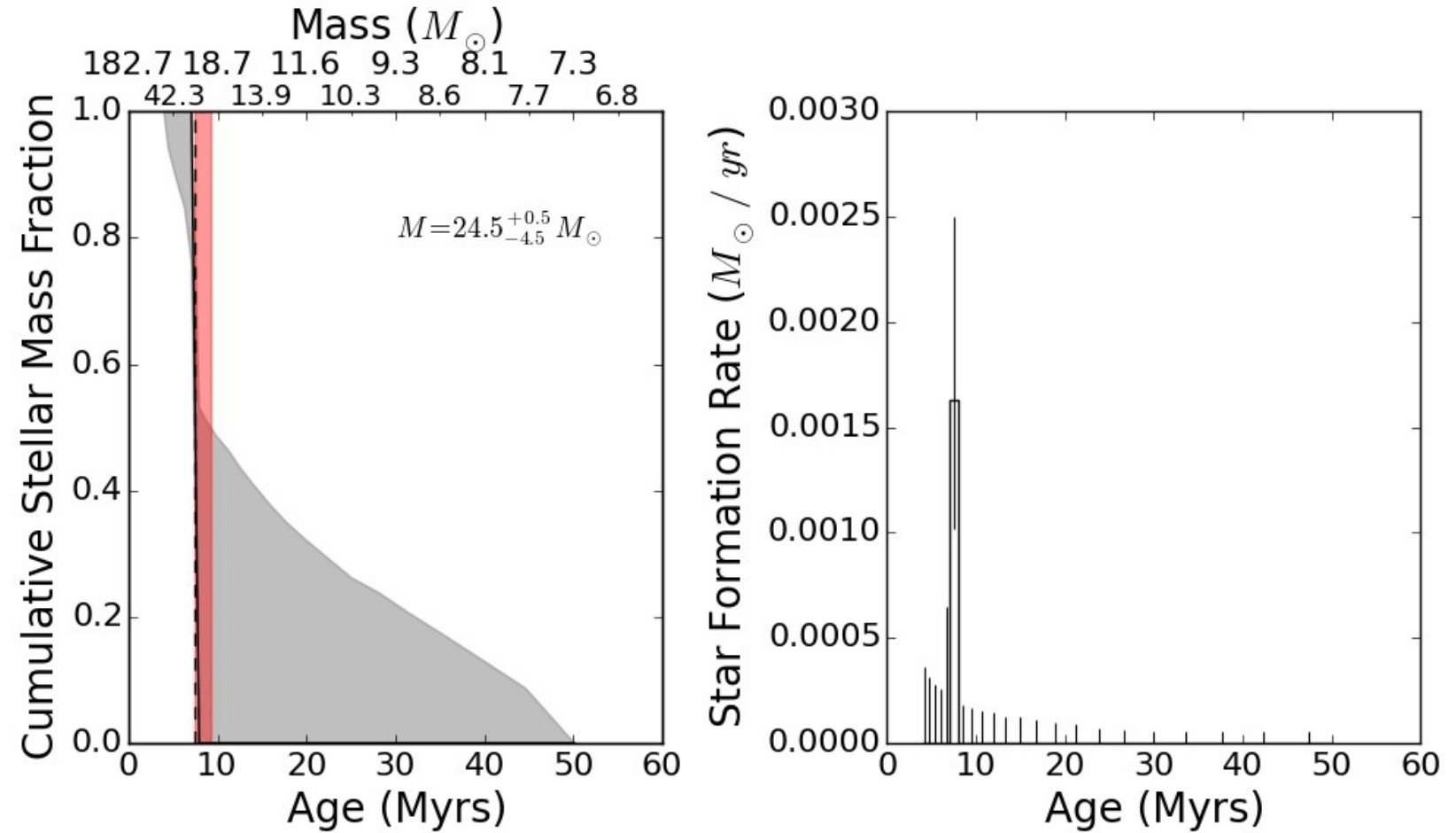}
    \caption{\small \\\hspace{\textwidth}\textbf{Left:} Cumulative stellar mass fraction vs. age and mass used to constrain the age of the SNR progenitor star. The dashed vertical line marks the most likely age of the population surrounding the SNR, and the gray area shows the uncertainties on the cumulative fraction at each age. The red shaded region shows the median age of the population and the uncertainty.    \\\hspace{\textwidth}\textbf{Right:} Star formation rate vs age from the fit, along with the uncertainties.  A clear peak at $8$~Myr is detected. This distribution results in the cumulative distribution and errors shown in left, where the fraction makes a rapid rise at $8$ Myr ago.}
    \label{fig:opt_mass}
\end{figure*}

\begin{figure*}[!p]
    \centering
	\includegraphics[width = \textwidth]{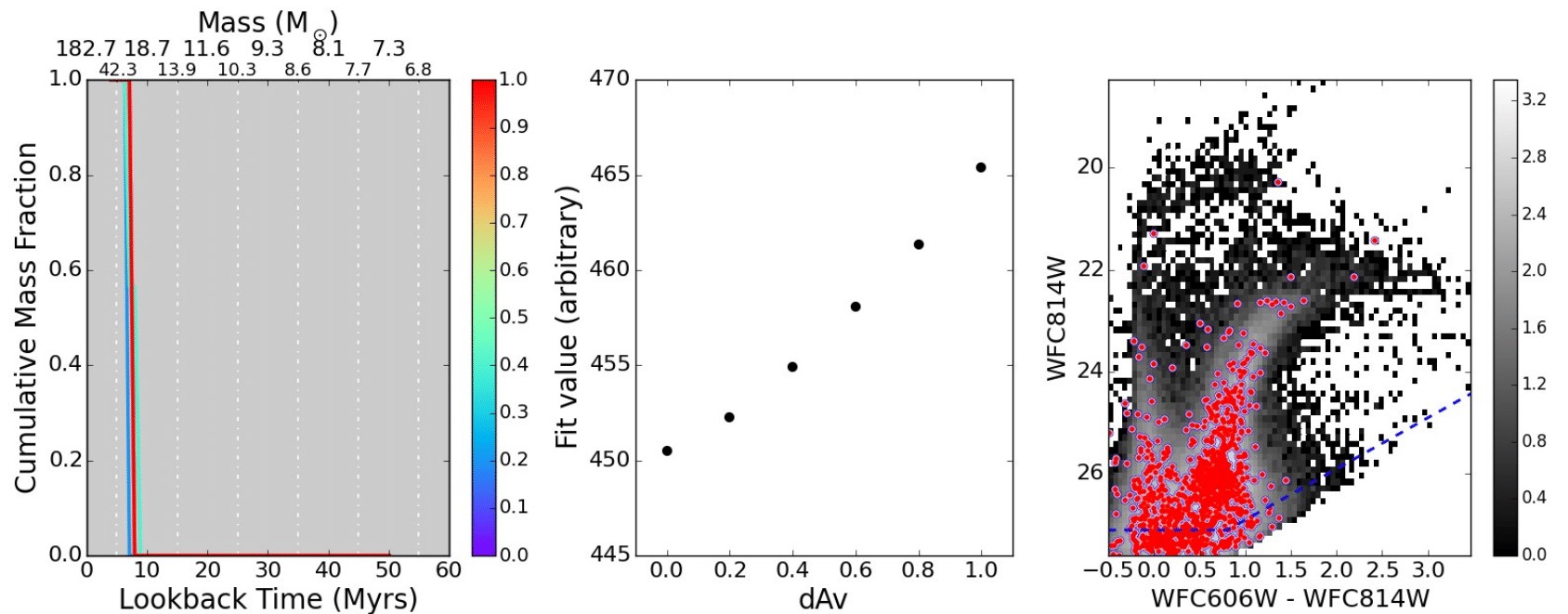}
    \caption{\small \\\hspace{\textwidth}\textbf{Left:} Cumulative Stellar Mass Fraction versus age (Lookback Time) for several different assumed dAv values.  The colors represent different fits, and the result appears insensitive to the choice of dAv.  \\\hspace{\textwidth}\textbf{Middle:} Plot depicting the fit values (lower is a better fit) based on Poisson maximum likelihood values \citep{Dolphin2000,Dolphin2012,Dolphin2013} for each assumed dAv.  The best-fit is for dAv$= 0$.  This value was adopted, and the final result is shown in Figure~\ref{fig:opt_mass}  \\\hspace{\textwidth}\textbf{Right:} Color Magnitude Diagram of the region around N$300$-S$26$ used to find the star formation history in \Cref{fig:opt_mass}.  The red points are the $1576$ stars within $50$ pc of the SNR center, and the grayscale shows the distribution of the remaining $376$,$000$ stars in the field, which were scaled by area and applied as a background sample during the fitting. The dashed line shows the $50$\% completeness limit.  The area above this line was included in the fit.}
    \label{fig:opt_mass_dav}
\end{figure*}

To determine the ages of the nearby stars, we begin with a CMD of the sample of $1576$ stars within $50$ pc of the SNR center. These are shown with the red points in Figure~\ref{fig:opt_mass_dav}.  The grayscale in the plot shows the remaining $376,000$ stars in the field, which were scaled by area and used as a background sample for the fitting.  The red plume of stars at WFC$606$W-WFC$814$W$\sim 1$ is the red giant branch, and it is made up of old ($>500$ Myr) stars.  The blue plume at WFC$606$W-WFC$814$W$\sim 0$ is the upper main sequence, and it consists of massive ($>3$ M$_{\odot}$) young ($<500$ Myr) stars.  There are many more old stars than young ones, but the strong upper main sequence presence at this location in NGC~$300$ suggests that our assumption of a massive progenitor is reasonable.  We then fit the CMD using the MATCH package \citep{Dolphin2012,Dolphin2013} to constrain the age distribution.  While MATCH returns the ages up to $13$ Gyr, we focus on the young component relevant to the SNR.

We show our results in \Cref{fig:opt_mass}, where the population clearly shows a strong peak at an age of $8$~Myr. We can use this age to infer the initial mass of the progenitor star assuming standard single star evolution. This age corresponds to the expected lifetime of a $25$~M$_{\odot}$ star. The uncertainties of our age distribution, as determined from the MATCH {\tt hybridmc} package \citep{Dolphin2012,Dolphin2013}, are shown by the red shading in Figure~\ref{fig:opt_mass}, and correspond to \SI{8 \pm 1}{\mega yr} and $25^{+1}_{-5}$ M$_{\odot}$, assuming the Padova stellar evolution models \citep{marigo2008,girardi2010}. This, therefore, suggests that this is one of the most massive SNR progenitors found in the Local Volume. The CMD associated with this population as well as the differential extinction can also be seen in \Cref{fig:opt_mass_dav} and they show that the result is not sensitive to the amount of differential extinction associated with this population. 

\subsection{X-Ray Spectra}
\label{sec:results_xray}



We simultaneously fit nine observations (total of $21$ different data sets due to the various detectors) from \textit{Chandra} and \textit{XMM-Newton}; each of the effective exposure times can be found in \Cref{tab:xray_obs}. We found that our best joint fit to the extracted spectra had a hydrogen column density of ($9.7^{+6.4}_{-4.8}$)$\times 10^{20}$ cm$^{-2}$ which is a slightly larger than the value for the foreground, $3.0 \times 10^{20}$ \SI{}{\per\square\cm}, obtained from COLDEN,\footnote{http://cxc.harvard.edu/toolkit/colden.jsp}, suggesting a small amount of extinction due to NGC $300$ as well as possibly from S$26$.

For each data set, we first fit a model to the extracted background data and found the normalization values for the sky and instrument modeled backgrounds. Then, we fit all of the source data with the modeled background data. The background models were all frozen to the values determined from their individual fits.

\begin{figure*}[!htb]
    \centering
	\includegraphics[width=\linewidth]{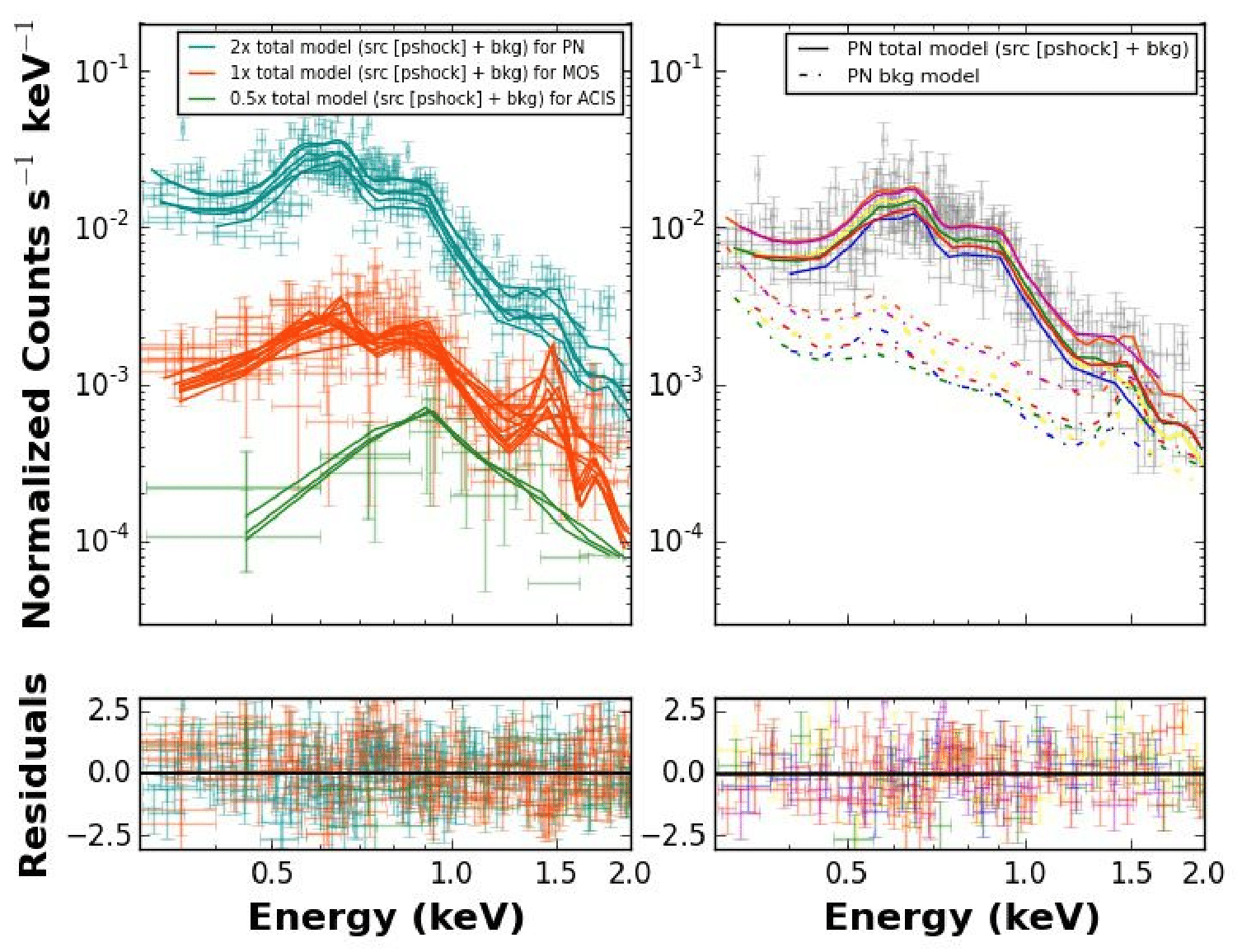}
    \caption{\small X-ray spectral fit using {\tt tbabs*(pshock)} and having the normalizations tied. The right-hand graph is the PN data with the six observations ($0112800101$, $0112800201$, $0305860301$, $0305860401$, $0791010101$, and $0791010301$) corresponding to blue, yellow, green, red, orange, and magenta, respectively.}
    \label{fig:xray_spectra}
\end{figure*}

We fit the source data to a model consisting of {\tt tbabs} and {\tt pshock} components using a distance of \SI{2}{\mega\parsec} and a metal abundance of $0.5$ (calculated using the metallicity gradient found from \cite{Gazak2015} and using a distance of $3.7$ kpc between S$26$ and the center of NGC $300$). The fitted values can be found in Table~\ref{tab:xray_fits} for having all of the normalizations tied together (see \Cref{fig:xray_spectra} for the plotted spectra). The fit quality for the tied normalizations was excellent showing robust cross-calibration of the extractions. We also fit the spectra with the normalizations not tied together, the normalizations tied together and extending the energy range to be between $0.3$ and $5.0$ keV, and the normalizations tied together but the temperatures allowed to vary from observation to observation and detector to detector (see \Cref{tab:xray_temp} for a breakdown of the varying temperatures), but found no significant change in the resulting parameter values (see \Cref{tab:xray_fits}). The values from the models in which normalizations were not tied together did result in significantly different parameters, but the fit was not significantly better and the tied normalizations are more physically plausible.

\begin{table*}[!p]
	\centering
    \begin{threeparttable}
      \resizebox{\textwidth}{!}{
      \begin{tabular}{cccc}
          \hline
          \hline
                                                         			  & {\tt tbabs*(pshock)}                & {\tt tbabs*(pshock)}                & {\tt tbabs*(pshock)}                 \\
          Parameter                                      			  & Normalizations Tied                & Normalizations Free                & Normalizations Tied                \\
                                                                      & ($90$\% Confidence)                &                                    & $0.3-5.0$ keV                 \\
          \hline
          $N_{H}$ ($10^{22}$ cm$^{-2}$)                  			  & ($9.7^{+6.4}_{-4.8}$)$\times 10^{-2}$  & ($9.7 \pm 2.1$)$\times 10^{-2}$ & $0.11 \pm 0.02$                     \\
          kT (keV)                                       			  & $0.77 \pm 0.13$                    & $0.77 \pm 0.06$                    & $0.73 \pm 0.05$                    \\
          <PShock Normalization>\tnotex{tnote:normps}    			  & ($5.4^{+2.0}_{-1.1}$)$\times 10^{-6}$  & ($5.5 \pm 0.45$)$\times 10^{-6}$  & ($5.7 \pm 1.1$)$\times 10^{-6}$  \\
          Photon Index                                                & ...                                & ...                                & ...                                \\ 
          <PowerLaw Normalization>\tnotex{tnote:normpl}               & ...                                & ...                                & ...                                \\ 
          Cash Statistic/dof                             			  & 7735/9025                         & 7716/9005                         & 15960/29915                        \\
          <Un-absorbed Luminosity> (erg s$^{-1}$)\tnotex{tnote:lumin} & $\sim 6.3 \times 10^{36}$ 			   & ...                                & ...                                \\
          \hline
          \hline
                                                                   	  & {\tt tbabs*(pshock)}                & {\tt tbabs*(pshock+powerlaw)}      & {\tt tbabs*(powerlaw)} \\
          Parameter                                      			  & Normalizations Tied                & Normalizations Free               & Normalizations Free \\
                                                                      &  Temperature Free                  &                                   & \\
          \hline
          $N_{H}$ ($10^{22}$ cm$^{-2}$)                  			  & ($9.4 \pm 1.9$)$\times 10^{-2}$    & $0.30 \pm 0.08$                   & $0.87 \pm 0.07$ \\
          kT (keV)                                       			  & $0.79 \pm 0.04$\tnotex{tnote:temp} & $0.53 \pm 0.09$                   & ... \\
          <PShock Normalization>\tnotex{tnote:normps}    			  & ($5.3 \pm 0.9$)$\times 10^{-6}$    & ($9.3 \pm 2.4$)$\times 10^{-6}$   & ... \\
          Photon Index                                                & ...                                & $5.3 \pm 0.6$                     & $8.4 \pm 0.5$ \\ 
          <PowerLaw Normalization>\tnotex{tnote:normpl}               & ...                                & ($2.1 \pm 0.9$)$\times 10^{-6}$  & ($2.2 \pm 0.2$)$\times 10^{-5}$ \\ 
          Cash Statistic/dof                             			  & 7725/9005                          & 7574/8983                         & 7755/9005 \\
          <Un-absorbed Luminosity> (erg s$^{-1}$)\tnotex{tnote:lumin} & ...                                & ...                               & ... \\
          \hline
      \end{tabular}
      }
      \begin{tablenotes}
      	\scriptsize
        \item\label{tnote:normps}Average normalization for pshock component of fit. For tied normalization fits, the average value is just the normalization value fitted. Defined as $(10^{-14}/(4\pi [D_{A}(1+z)]^{2})) \int n_{e}n_{H} dV$ where $D_{A}$ is the distance to the source in cm, $z$ is the redshift, $n_{e}$ is the electron number density in cm$^{-3}$, and $n_{H}$ is the hydrogen number density in cm$^{-3}$.
        \item\label{tnote:normpl}Average normalization for powerlaw component of fit. Defined as photons/keV/cm$^{2}$/s at $1$ keV.
        \item\label{tnote:temp}Average temperature over the various observations and various detectors. For a more detailed view of the temperature values for the different observations and detectors for the {\tt tbabs*(pshock)}, normalizations tied, temperature free fit see \Cref{tab:xray_temp}.
        \item\label{tnote:lumin}Average un-absorbed luminosity over the nine different observations (six \textit{XMM-Newton} and three \textit{Chandra}) between the $0.3-2.0$ keV energy range.
        \par
      \end{tablenotes}
    \end{threeparttable}
    \caption{\small X-ray spectral fits to the source data based on various different fit parameters. Best-fit spectral fit corresponds to the {\tt tbabs*(pshock)} with the normalizations tied. The standard practice for these fits are to list the $1\sigma$ uncertainties for the values and to have the energy levels be between $0.3-2.0$ keV. This is the practice used unless otherwise stated.}
	\label{tab:xray_fits}
\end{table*}

\begin{table*}[!p]
	\centering
    \begin{tabular}{cccc}
    \hline
    \hline
   	Obs ID     & Detector               & Date         & kT              \\ 
               & (With Filter)          & (yyyy mm dd) & (keV)           \\
    \hline
    12238      & ACIS-I VFAINT          & 2010 Sep 24  & $0.71 \pm 0.15$ \\
    16028      & ACIS-I FAINT           & 2014 May 16  & $0.78 \pm 0.13$ \\
    16029      & ACIS-I FAINT           & 2014 Nov 17  & $0.66 \pm 0.11$ \\
    \hline
               & EPIC-MOS$1$ Medium     &              & $0.93 \pm 0.21$ \\
    0112800101 & EPIC-MOS$2$ Medium     & 2001 Jan 2   & $0.80 \pm 0.17$ \\
               & EPIC-PN Medium         &              & $0.86 \pm 0.14$ \\
    \hline
               & EPIC-MOS$1$ Medium     &              & $0.75 \pm 0.22$ \\
    0112800201 & EPIC-MOS$2$ Medium     & 2000 Dec 27  & $0.85 \pm 0.22$ \\
               & EPIC-PN Medium         &              & $1.0 \pm 0.2$ \\
    \hline
               & EPIC-MOS$1$ Medium     &              & $0.55 \pm 0.11$ \\
    0305860301 & EPIC-MOS$2$ Medium     & 2005 Nov 25  & $0.86 \pm 0.18$ \\
               & EPIC-PN Medium         &              & $0.80 \pm 0.12$ \\
    \hline
               & EPIC-MOS$1$ Medium     &              & $0.74 \pm 0.22$ \\
    0305860401 & EPIC-MOS$2$ Medium     & 2005 May 22  & $0.90 \pm 0.35$ \\
               & EPIC-PN Medium         &              & $0.87 \pm 0.17$ \\
    \hline
               & EPIC-MOS$1$ Thin$1$    &              & $0.68 \pm 0.09$ \\
    0791010101 & EPIC-MOS$2$ Medium     & 2016 Dec 17  & $0.76 \pm 0.11$ \\
               & EPIC-PN Medium         &              & $0.76 \pm 0.08$ \\
    \hline
               & EPIC-MOS$1$ Thin$1$    &              & $0.66 \pm 0.12$ \\
    0791010301 & EPIC-MOS$2$ Medium     & 2016 Dec 19  & $0.85 \pm 0.17$ \\
               & EPIC-PN Medium         &              & $0.69 \pm 0.10$ \\
    \hline
    \end{tabular}
    \caption{\small The values for the temperature with $1\sigma$ uncertainties in the {\tt tbabs*(pshock)}, normalizations tied, temperature free fit.}
    \label{tab:xray_temp}
\end{table*}

The best-fit temperature was $0.77 \pm 0.13$ keV which is close to the typically measured temperature of $0.7\ \kilo\eV$ for nearby SNRs \citep{BlairLong2010,Maggi2016}. This temperature could be indicative of a pulsar heating up the surrounding gas or due to the SNR being young and expanding into a dense ambient medium. To test if there was a pulsar heating up the gas, we fit the source data to a model with {\tt tbabs}, {\tt pshock}, and {\tt powerlaw} components. However, the results did not reveal any significant detection of a hard component to the spectrum (see \Cref{tab:xray_fits}). We also ran a fit that only had {\tt tbabs} and {\tt powerlaw} components, but also found no evidence of a hard component to the spectrum (see \Cref{tab:xray_fits}).

Previous observations of S$26$ also found a best-fit for the hydrogen column density and the temperature. \cite{Read1997} fit the data to a thermal Bremsstrahlung model and found the hydrogen column density to be $6.98^{+17.6}_{-3.2} \times 10^{20}$ cm$^{-s}$ and the temperature to be $0.51^{+0.45}_{-0.29}$ keV. Both of our own best-fit values for the hydrogen column density and the temperature fall within the errors found by \cite{Read1997}, which further supports our fits.


We also estimated a luminosity over the $0.3-2.0$ keV energy band using the XSPEC command {\tt lumin} to be $\sim 3.9 \times 10^{36}$ erg s$^{-1}$, using a distance of \SI{2}{\mega\parsec}, and an unabsorbed luminosity of $\sim 6.3 \times 10^{36}$ erg s$^{-1}$.

The wings of the PSF of the nearest neighbor may result in a small amount of contamination in some of the extracted spectra (see \Cref{fig:xray_extraction}). In order to investigate the effects of such contamination, we ran a separate set of extractions where we masked any area within the $23.9$ arcsec radius around this neighboring source. This alteration results in a significant loss of SNR counts, as the SNR light dominates in these locations; however, it is important to determine the maximum effects of the neighboring source.

We fit this masked out source region to a {\tt tbabs*(pshock)} model and in order to better constrain this fit, we allowed the various normalizations from observation to observation and detector to detector to be varied individually since the relative amount of counts removed from each data set were too different from one another to justify the normalizations to be tied together. This model resulted in a hydrogen column density of ($7.0 \pm 2.0$)$\times 10^{20}$ cm$^{-2}$, a temperature of $0.75 \pm 0.05$ keV, an average normalization of ($4.4 \pm 11.3$)$\times 10^{-5}$ cm$^{-5}$, a luminosity of $\sim 3.5 \times 10^{37}$ erg s$^{-1}$, an unabsorbed luminosity of $\sim 5.1 \times 10^{37}$ erg s$^{-1}$, and a Cash Statistic/dof of $4855/4799$ with these errors corresponding to the $1\sigma$ limit. These values are all within the errors derived for our fit to the full extraction area stated above, which suggests that the nearby source was not significantly affecting the fit.

As an additional check on our fitting, we also performed a very simple fit to only the PN detector data for observation $0791010101$ since it had the most counts out of all of the observations and detectors and was one of the more on-axis observations that we had. For the total source extraction region and a {\tt tbabs*(pshock)} model, the hydrogen column density was ($10 \pm 5$)$\times 10^{20}$ cm$^{-2}$, the temperature was $0.73 \pm 0.11$ keV, the normalization was ($5.5 \pm 2.2$)$\times 10^{-6}$ cm$^{-5}$, and the Cash Statistic/dof was $732/518$. For the extraction region altered to remove any extended wings from the neighboring source and a {\tt tbabs*(pshock)} model, the hydrogen column density was ($12 \pm 5$)$\times 10^{20}$ cm$^{-2}$, the temperature was $0.63 \pm 0.10$ keV, the normalization was ($4.3 \pm 1.8$)$\times 10^{-5}$ cm$^{-5}$, and the Cast Statistic/dof was $705/499$. The temperatures and absorption parameters reported for each of these fits are equivalent within their uncertainties, and they are consistent with those for the full data set (see \Cref{tab:xray_fits}), again suggesting that the wings from the neighboring source are not strongly affecting the fits to the SNR spectrum.

\subsection{Radio Properties of N300-S26}
\label{sec:results_radio}

We used the previous observations by \cite{Pannuti2000} and \cite{Payne2004} and the newly found radius from \textit{HST} to evaluate the integrated radio flux density of the SNR (from this we infer the total radio luminosity, $L$), the
minimum total energy, $E_{\rm{min}}$, corresponding to the energy stored within the SNR in the form of relativistic particles, and the corresponding minimum magnetic field, $B_{\rm{min}}$.

Following \cite{Pacholczyk1970}, where equipartition is assumed (magnetic energy is equal to the total particle energy), the radio flux density, $S_{\rm{\nu}}$, may be expressed in the form of a power law as a function of frequency as 
\begin{equation}
S_{\rm{\nu}} = \beta \nu^{-\alpha}
\end{equation}
where $\beta$ is a constant. From this definition, the integrated flux density, $S$,
over the frequency range $\nu_{\rm{1}}$ to $\nu_{\rm{2}}$ is 
\begin{equation}
S = \int_{\nu_1}^{\nu_2} S_{\rm{\nu}} d\nu
\end{equation}
where $\nu_{\rm{1}}$ and $\nu_{\rm{2}}$ are $10^8$ Hz and $10^{11}$ Hz, respectively. The integrated radio luminosity, $L$, of S$26$ is
\begin{equation}
L = 4 \pi d^2 S = 4 \pi d^2 \int_{10^8~Hz}^{10^{11}~Hz} \beta \nu^{-\alpha}~d\nu,
\end{equation}
where $d$ is the distance to the SNR. 

The measured flux density at \SI{1.448}{\giga\hertz} and the radio spectral index for S$26$ was $0.22$ mJy and a lower limit to $\alpha$ of $0.65$, respectively \citep{Pannuti2000,Payne2004}. This results in a value for $\beta$ of $1.9 \times 10^{-21}$ erg cm$^{-2}$ s$^{-1}$ Hz$^{-1/4}$ for the values of $\alpha$ and flux density stated. Using these values and our adopted distance to NGC $300$, we calculate a lower limit for the radio luminosity $L$ of $\sim 1.7\times 10^{34}$ erg s$^{-1}$ over the frequency range between $10^{8}$ Hz and $10^{11}$ Hz.

Next, we calculate the upper limit values for both $B_{\rm{min}}$ and $E_{\rm{min}}$ following \cite{Pacholczyk1970}:
\begin{eqnarray*}
B_{\rm{min}} & = & (4/5)^{5/7} c_{\rm{12}}^{2/7} (1+\xi)^{2/7}  f^{-2/7} r^{-6/7} L^{2/7} \\
E_{\rm{min}} & = & c_{13} (1+\xi)^{4/7} f^{3/7} r^{9/7} L^{4/7}
\end{eqnarray*}
where $\xi$ is the ratio of the energies of the high mass ions and the relativistic electrons within the SNR (we assume a value of $40$, which is consistent with measured values for this ratio at the top of the Earth's atmosphere), $f$ is a filling factor corresponding to the volume of the SNR that is filled by the magnetic fields and relativistic particles (we assume a value of $0.25$ that is consistent with values for an SNR in the Sedov stage of evolution; \cite{Duric1995,Pannuti2000PhD}), and $r$ is the radius of S$26$ (an mean value of \SI{7.05}{\parsec} from our measured ellipse). Finally, the upper limits for $c_{\rm{12}}$ and $c_{\rm{13}}$ are
\begin{eqnarray*}
c_{\rm{12}} & = & \left( \frac{c_{\rm{1}}^{1/2}}{c_{\rm{2}}} \right) \left( \frac{2\alpha - 2}{2\alpha - 1} \right) 
\left(\frac{\nu_{\rm{1}}^{\frac{1-2\alpha}{2}}-\nu_{\rm{2}}^{\frac{1-2\alpha}{2}}}{\nu_{\rm{1}}^{1-\alpha} - \nu_{\rm{2}}^{1-\alpha}} \right) \\
c_{\rm{13}} & = & 0.921 c_{\rm{12}}^{4/7}    
\end{eqnarray*}
and in turn $c_1$ and $c_2$ are
\begin{eqnarray*}
c_{\rm{1}} & = & \frac{3e}{4 \pi m_{\rm{e}}^3 c^5} = 6.27 \times 10^{18} \\
c_{\rm{2}} & = & \frac{2e^4}{3 m_e^4 c^7} = 2.37 \times 10^{-3} \\
\end{eqnarray*}
where $e$ is the charge of the electron, $m_{e}$ is the mass of the electron, and $c$ is the speed of light in a vacuum; therefore, $c_{\rm{1}}$ has units of statCoulomb s$^{5}$ g$^{-3}$ cm$^{-5}$ and $c_{\rm{2}}$ has units of statCoulomb$^{4}$ s$^{7}$ g$^{-4}$ cm$^{-7}$. These equations yield values of $0.067$ mG and $1.5\times 10^{49}$ erg for the maximum values of $B_{\rm{min}}$ and $E_{\rm{min}}$, respectively.

To obtain context for the measured values for $L$, $B_{min}$, and $E_{min}$ for
S$26$, we examined published values for these properties for Galactic and extragalactic SNRs \citep{Duric1995,Pannuti2000PhD,Lacey2001,Reynolds2012}. We also computed the values for these properties for a sample of $23$ Galactic SNRs from \cite{Green2017}\footnote{See http://www.mrao.cam.ac.uk/surveys/snrs/}. To generate this sample, we chose SNRs with clear shell-like radio morphologies, which also had robust measurements of spectral indices, flux densities at $1$ GHz, angular extents, and distances. Known historical SNRs like Cassiopeia A and SN $1006$, which may still be in the free-expansion stage of evolution rather than the Sedov stage, were excluded from our study. For our sample, the median values of $L$, $B_{min}$, and $E_{min}$ were $8.30 \times 10^{33}$ erg s$^{-1}$, $0.027$ mG, and $3.35 \times 10^{49}$ erg, respectively; the standard deviations of the values for $L$, $B_{min}$, and $E_{min}$ were $5.48 \times 10^{34}$ erg s$^{-1}$, $0.028$ mG, and $6.69 \times 10^{49}$ erg, respectively.

Our computed values for $L$ and $E_{min}$ for S$26$ are both within one standard deviation of the computed median values for our sample of Galactic SNRs. Furthermore, they are both an order of magnitude less than the computed range of values for these parameters presented by \cite{Lacey2001} for the sample of candidate radio SNRs in the galaxy NGC $6946$. This discrepancy may be explained in part by the limiting sensitivity attained by the survey conducted by \cite{Lacey2001}, which could not detect candidate radio SNRs as faint as S$26$ at the distance of NGC $6946$ ($5.9$ Mpc; \cite{Karachentsev2000}). In addition, the elevated star formation rate of NGC $6946$ relative to NGC $300$ produces indirectly a significant population of luminous resident candidate radio SNRs with proportionally higher values of $L$ and $E_{min}$. The radio luminosity of S$26$ is also fainter than the calculated radio luminosities for candidate radio SNRs in M$33$ by \cite{Duric1995} who assumed a distance of $840$ kpc to that galaxy.

Regarding $B_{min}$, while our computed value for this property falls within the range of measured values for Galactic SNRs from \cite{Reynolds2012} it is more than one standard deviation greater than the median value for $B_{min}$ for the Galactic SNRs in our sample. It is also an order of magnitude less than the computed range of values computed by \cite{Lacey2001} for the candidate radio SNRs in NGC $6946$. We attribute the difference between our computed value for S$26$'s $B_{min}$ to the computed values for $B_{min}$ by \cite{Lacey2001} to limiting sensitivity of the radio observations analyzed in that work (similar to the cases of the discrepancies with computed values of $L$ and $E_{min}$) but we cannot readily explain why the value
of $B_{min}$ is significantly greater than the median value found for our sample of Galactic SNRs. New radio observations of S$26$ with enhanced sensitivity that can more clearly determine its spectral index of this source may help address this question.

\section{Discussion}
\label{sec:discussion}

The temperature from our X-ray spectral fits and the radius of the source from \textit{HST} allow us to estimate how long ago the supernova (SN) occurred. From Equation $2$ of \cite{Hughes1998}, this age is
\begin{equation}
\label{eq:ageHughes}
t\ (yr) = 10^{3} \bigg(\frac{kT}{1 \kilo\electronvolt}\bigg)^{-1/2}\bigg(\frac{\theta_{R}}{10"}\bigg)\bigg(\frac{D}{50 \kilo\parsec}\bigg)
\end{equation}
where $kT$ is the temperature derived from the best-fit spectral model in keV, $\theta_{R}$ is the angular size in arcsec, and $D$ is the distance to the source in kiloparsecs. We calculated that the SN associated with S$26$ occurred ($3.3^{+0.7}_{-0.6}$)$\times 10^{3}$ yr ago for S$26$. 



We also calculate the electron number density, $n_{e}$, and hydrogen number density, $n_{H}$, from the normalization found in the spectral fits. Using 
\begin{equation}
normalization = \frac{10^{-14}}{4\pi [D_{A}(1+z)]^{2}} \int n_{e}n_{H} dV
\end{equation}
where $D_{A}$ is the distance to the source in centimeters and $z$ is the redshift---$6 \times 10^{24}$ cm and $4 \times 10^{-4}$, respectively. We assume that $n_{e} = 1.2 n_{H}$ and that $n_{H}$ is constant throughout the SNR and has units of cm$^{-3}$. The $\int dV$ is just the spherical volume of the SNR using a radius of $7.1$ pc (the average radius from the semimajor and semiminor axis). The normalization of ($5.4 \pm 1.0$)$\times 10^{-6}$ cm$^{-5}$ results in a $n_{H}$ of $2.2^{+0.9}_{-0.6}$ cm$^{-3}$ and a $n_{e}$ of $2.6^{+1.1}_{-0.7}$ cm$^{-3}$, again consistent with our young age and high density hypothesis. It is worth noting that we have assumed a constant $n_{H}$ for this SNR due to there not being enough counts to have a model with more than one hydrogen density value. There could very well be a variable $n_{H}$ across the SNR, but with our limited photon statistics, we were only able to simplify our model to derive a constant $n_{H}$ value.

We also derive the corresponding pressure of the SNR using
\begin{equation}
P/k\ (K\ cm^{-3}) = 2 n_{e} T
\end{equation}
where $n_{e}$ is the electron number density in cm$^{-3}$ with a factor of $2$ to represent the total number of particles, both electrons and protons, $T$ is the temperature of the SNR in K, and \textit{k} is Boltzmann's Constant. With the $n_{e}$ derived above and the temperature associated with $0.77 \pm 0.05$ keV (error at $1\sigma$ level), which corresponds to ($8.9 \pm 0.5$)$\times 10^{6}$ K, results in a \textit{P/k} value of ($4.7^{+2.3}_{-1.5}$)$\times 10^{7}$ K cm$^{-3}$ which correspond to a pressure of ($6.5^{+3.2}_{-2.1}$)$\times 10^{-10}$ Pa. This value can help us compare S$26$ to other SNRs by looking at the pressure due to the shock wave of both of these SNe. It is worth noting that this is a rough approximation because the ideal gas law is not perfectly applicable for this SNR because the ionization timescale is of the order of $10^{11}$, which is outside of collisional ionization equilibrium.

We have also calculated the amount of mass swept up by the SNR using
\begin{equation}
M_{X} = f m_{H} n_{H} V
\end{equation}
where $f$ is the volume filling factor of the X-ray emitting plasma (assumed to be $1$ for simplicity), $m_{H}$ is the mass of a hydrogen atom, $n_{H}$ is the hydrogen number density calculated from the normalization above, and $V$ is the volume of the SNR (the ellipse from optical emission is assumed to be edge on for simplicity). This resulted in a $M_{X}$ of $82^{+100}_{-47}$ \(M_{\odot}\). This value can help us compare S$26$ to other SNRs by looking at the amount of matter swept up by the shock wave of both of these SNe as they propagate through space.


In order to get an estimate on the shock velocity of S$26$, we used conservation of energy and assumed that the deceleration of the material was constant after the second phase evolution of the SNR began (when $1$ \(M_{\odot}\) of matter was swept up). Doing so resulted in a transition radius of $1.6^{+0.3}_{-0.1}$ pc using a density $\rho = m_{H}n_{H} = $($3.7^{+1.5}_{-1.0}$)$\times 10^{-24}$ g cm$^{-3}$ where $m_{H}$ is the mass of hydrogen in g and $n_{H}$ is the value derived above. We then assumed that the initial velocity at the beginning of the second phase was $3700$ km s$^{-1}$, which was taken from \cite{Michael2003}. This resulted in a shock velocity of $411^{+275}_{-122}$ km s$^{-1}$ and an age for the SNR of ($2.6^{+0.7}_{-0.8}$)$\times 10^{3}$ yr. This age overlaps the age derived from Equation \ref{eq:ageHughes}, which helps support this shock velocity.

To put S$26$ into context, we compare the values derived with Galactic SNRs. One such example of a Galactic SNR that has a similar temperature and size is G$311.5$--$0.3$, which was examined in detail in \cite{Pannuti2014,Pannuti2017}. G$311.5$--$0.3$ is at a distance of $12.5$ kpc, has a temperature of $0.68^{+0.20}_{-0.24}$ keV, and a radius of \SI{9}{\parsec}; very similar to S$26$. \cite{Pannuti2014}, derived many quantities from the X-ray spectral fit to G$311.5$--$0.3$---a shock velocity of $38.7$ km s$^{-1}$, a remnant age of ($2.5-4.2$)$\times 10^{4}$ yr, a $n_{e}$ of $0.20$ cm$^{-3}$, a $M_{X}$ of 21.4 \(M_{\odot}\), a \textit{P/k} ratio of \SI{3.18e6}{\kelvin\per\cubic\cm}, and an unabsorbed luminosity of $4.09 \times 10^{34}$ erg s$^{-1}$. It is worth noting that \cite{Pannuti2014} did not report any uncertainties for these values.

It is somewhat surprising that G$311.5$--$0.3$ has a similar size and temperature yet is so much older and has such a lower shock velocity than S$26$. Some of these differences between S$26$ and G$311.5$--$0.3$ are likely related to the vast difference in X-ray luminosity. This factor of $1000$ in luminosity is consistent with a younger age and a denser medium. Naively, the ages of both of these SNRs should be similar if the remnant age is proportional to the radius and inversely proportional to the shock velocity ($v_{s} \sim \sqrt{kT}$). However, the shock velocity in S$26$ appears to be much higher leading to the large size at the relatively young age. Since both of these SNRs have similar radii and temperatures, one might expect that they would have similar ages. However, the age of S$26$ is an order of magnitude smaller. We note that the age for G$311.5$--$0.3$ is likely better constrained than for S$26$ because its shock velocity was measured from CO observations by \cite{Andersen2011} and \cite{Hewitt2009}; however, our measurements suggest very little chance that S$26$ is as old as G$311.5$--$0.3$, which is consistent with its much higher X-ray luminosity.


An explanation for the similar sizes and temperature despite very different ages could be attributable to the differences in the hydrogen number density. For S$26$, the $n_{H}$ is approximately an order of magnitude larger than G$311.5$--$0.3$. This higher number density could result in a higher shock velocity in S$26$ explaining the younger age for the SNR. The higher number density in S$26$ also explains why the amount of mass swept up is significantly higher, as well as the fact that the number density results in the \textit{P/k} ratio that is an order of magnitude higher for S$26$.

All of these differences support our hypothesis that the fitted temperature for S$26$ is due to this source being a young SNR that is expanding into a dense medium. It thus appears that the higher number density medium has played a large role in making S$26$ appear to be young even though the temperature and size are similar to G$311.5$--$0.3$.

\section{Conclusions}
\label{sec:conclusion}

From multiwavelength observations of S$26$ using data from \textit{Chandra}, \textit{XMM-Newton}, \textit{HST}, \textit{VLA}, and \textit{ATCA} we have constrained several of this SNR's physical properties. We have simultaneously fitted three different \textit{Chandra} observations and six \textit{XMM-Newton} observations to constrain the properties derived via X-ray observations as well as analyzing this SNR in the optical and radio wavelengths to get a more complete picture of this source.


We have measured a precise optical size for the SNR from \textit{HST} which shows this SNR was much smaller than previous, ground-based measurements indicated---it has a semimajor axis of $7.5 \pm 1.0\ \parsec$ and semiminor axis of $6.7^{+1.2}_{-1.4}\ \parsec$. From X-ray spectral fitting, we estimated an unabsorbed luminosity, temperature, and intervening hydrogen column density---$\sim 6.3 \times 10^{36}$ erg s$^{-1}$, $0.77 \pm 0.13$ keV, and ($9.7^{+6.4}_{-4.8}$)$\times 10^{20}$ cm$^{-2}$, respectively. From radio observations, we measured the radio luminosity over the $10^{8}-10^{11}$ Hz frequency range and found the maximum values for $B_{min}$ and $E_{min}$ that are consistent with SNRs that had a high mass progenitor star---$1.7\times 10^{34}$ erg s$^{-1}$, $0.067$ mG, and $1.5\times 10^{49}$ erg, respectively. The age of the surrounding population of stars also suggests that the SNR had a relatively high mass progenitor---\SI{8 \pm 1}{\mega yr} and $25^{+1}_{-5}$ M$_{\odot}$. Finally, we compared our extragalactic SNR to a similar Galactic SNR; namely, G$311.5$--$0.3$. S$26$ appears to be a younger and significantly more energetic SNR that is expanding into a denser medium than its Galactic analog---($3.3^{+0.7}_{-0.6}$)$\times 10^{3}$ yr old and $2.2^{+0.9}_{-0.6}$ cm$^{-3}$.


\section*{Acknowledgements}

We would like to thank the anonymous referee for all of their great suggestions and comments because they greatly improved the quality of our paper.

T.G.P. would like to thank Christina Lacey and Jeffrey Payne for many useful discussions about the radio properties of 
extragalactic supernova remnants and radio observations of NGC $300$.

We would like to thank the Mary Gates Fellowship for support during this project. 

Support for this work was also provided by the National Aeronautics and Space Administration through \textit{Chandra} Award Number GO$4$-$15088$X issued by the \textit{Chandra X-ray Observatory Center}, which is operated by the Smithsonian Astrophysical Observatory for and on behalf of the National Aeronautics and Space Administration under contract NAS$8$-$03060$. 

Support for this work was also provided by National Aeronautics and Space Administration through \textit{Hubble Space Telescope} Award Number GO-$13515$ from the Space Telescope Science Institute, which is operated by AURA, Inc., under NASA contract NAS $5$-$26555$.




\bibliographystyle{apj}

\begin{thebibliography}{99}
\expandafter\ifx\csname natexlab\endcsname\relax\def\natexlab#1{#1}\fi

\bibitem[{{Andersen} {et~al.}(2011){Andersen}, {Rho}, {Reach}, {Hewitt}, \&
  {Bernard}}]{Andersen2011}
{Andersen}, M., {Rho}, J., {Reach}, W.~T., {Hewitt}, J.~W., \& {Bernard}, J.~P.
  2011, \apj, 742, 7

\bibitem[{{Badenes} {et~al.}(2009){Badenes}, {Harris}, {Zaritsky}, \&
  {Prieto}}]{Badenes2009}
{Badenes}, C., {Harris}, J., {Zaritsky}, D., \& {Prieto}, J.~L. 2009, \apj,
  700, 727

\bibitem[{{Blair} \& {Long}(1997)}]{BlairLong1997}
{Blair}, W.~P., \& {Long}, K.~S. 1997, \apjs, 108, 261

\bibitem[{{Blair} {et~al.}(2012){Blair}, {Winkler}, \& {Long}}]{Blair2012}
{Blair}, W.~P., {Winkler}, P.~F., \& {Long}, K.~S. 2012, \apjs, 203, 8

\bibitem[{{Bozzetto} {et~al.}(2017){Bozzetto}, {Filipovi{\'c}}, {Vukoti{\'c}},
  {Pavlovi{\'c}}, {Uro{\v s}evi{\'c}}, {Kavanagh}, {Arbutina}, {Maggi},
  {Sasaki}, {Haberl}, {Crawford}, {Roper}, {Grieve}, \&
  {Points}}]{Bozzetto2017}
{Bozzetto}, L.~M., {Filipovi{\'c}}, M.~D., {Vukoti{\'c}}, B., {et~al.} 2017,
  \apjs, 230, 2

\bibitem[{{Carpano} {et~al.}(2005){Carpano}, {Wilms}, {Schirmer}, \&
  {Kendziorra}}]{Carpano2005}
{Carpano}, S., {Wilms}, J., {Schirmer}, M., \& {Kendziorra}, E. 2005, \aap,
  443, 103
  
\bibitem[Cash(1979)]{Cash1979} Cash, W.\ 1979, \apj, 228, 939.

\bibitem[{{Dalcanton} {et~al.}(2009){Dalcanton}, {Williams}, {Seth}, {Dolphin},
  {Holtzman}, {Rosema}, {Skillman}, {Cole}, {Girardi}, {Gogarten},
  {Karachentsev}, {Olsen}, {Weisz}, {Christensen}, {Freeman}, {Gilbert},
  {Gallart}, {Harris}, {Hodge}, {de Jong}, {Karachentseva}, {Mateo}, {Stetson},
  {Tavarez}, {Zaritsky}, {Governato}, \& {Quinn}}]{Dalcanton2009}
{Dalcanton}, J.~J., {Williams}, B.~F., {Seth}, A.~C., {et~al.} 2009, \apjs,
  183, 67

\bibitem[{{Desai} {et~al.}(2010){Desai}, {Chu}, {Gruendl}, {Dluger}, {Katz},
  {Wong}, {Chen}, {Looney}, {Hughes}, {Muller}, {Ott}, \& {Pineda}}]{Desai2010}
{Desai}, K.~M., {Chu}, Y.-H., {Gruendl}, R.~A., {et~al.} 2010, \aj, 140, 584

\bibitem[{{Dodorico} {et~al.}(1980){Dodorico}, {Dopita}, \&
  {Benvenuti}}]{DOdorico1980}
{Dodorico}, S., {Dopita}, M.~A., \& {Benvenuti}, P. 1980, \aaps, 40, 67

\bibitem[{{Dolphin}(2000)}]{Dolphin2000}
{Dolphin}, A.~E. 2000, \pasp, 112, 1383

\bibitem[{{Dolphin}(2012)}]{Dolphin2012}
---. 2012, \apj, 751, 60

\bibitem[{{Dolphin}(2013)}]{Dolphin2013}
---. 2013, \apj, 775, 76

\bibitem[{{Duric}(1995)}]{Duric1995} Duric, N., Gordon, S.~M., Goss, W.~M., et al.\ 1995, \apj, 445, 173.

\bibitem[{{Freedman} {et~al.}(1992){Freedman}, {Madore}, {Hawley}, {Horowitz},
  {Mould}, {Navarrete}, \& {Sallmen}}]{Freedman1992}
{Freedman}, W.~L., {Madore}, B.~F., {Hawley}, S.~L., {et~al.} 1992, \apj, 396,
  80

\bibitem[Garofali et al.(2017)]{Garofali2017} Garofali, K., Williams, B.~F., Plucinsky, P.~P., et al.\ 2017, \mnras, 472, 308

\bibitem[{{Gazak} {et~al.}(2015){Gazak}, {Kudritzki}, {Evans}, {Patrick},
  {Davies}, {Bergemann}, {Plez}, {Bresolin}, {Bender}, {Wegner}, {Bonanos}, \&
  {Williams}}]{Gazak2015}
{Gazak}, J.~Z., {Kudritzki}, R., {Evans}, C., {et~al.} 2015, \apj, 805, 182

\bibitem[{{Girardi} {et~al.}(2010){Girardi}, {Williams}, {Gilbert},
  {Rosenfield}, {Dalcanton}, {Marigo}, {Boyer}, {Dolphin}, {Weisz},
  {Melbourne}, {Olsen}, {Seth}, \& {Skillman}}]{girardi2010}
{Girardi}, L., {Williams}, B.~F., {Gilbert}, K.~M., {et~al.} 2010, \apj, 724,
  1030

\bibitem[{{Gogarten} {et~al.}(2009){Gogarten}, {Dalcanton}, {Williams}, {Seth},
  {Dolphin}, {Weisz}, {Skillman}, {Holtzman}, {Cole}, {Girardi}, {de Jong},
  {Karachentsev}, {Olsen}, \& {Rosema}}]{Gogarten2009}
{Gogarten}, S.~M., {Dalcanton}, J.~J., {Williams}, B.~F., {et~al.} 2009, \apj,
  691, 115
  
\bibitem[{Green}(2017)]{Green2017} {Green, D.A.} 2017, `A Catalogue of Galactic Supernova Remnants (2017 June Version)', Cavendish Laboratory, Cambridge, United Kingdom

\bibitem[{{Hewitt} {et~al.}(2009){Hewitt}, {Rho}, {Andersen}, \&
  {Reach}}]{Hewitt2009}
{Hewitt}, J.~W., {Rho}, J., {Andersen}, M., \& {Reach}, W.~T. 2009, \apj, 694,
  1266

\bibitem[{{Hughes} {et~al.}(1998){Hughes}, {Hayashi}, \& {Koyama}}]{Hughes1998}
{Hughes}, J.~P., {Hayashi}, I., \& {Koyama}, K. 1998, \apj, 505, 732

\bibitem[{{Jennings} {et~al.}(2012){Jennings}, {Williams}, {Murphy},
  {Dalcanton}, {Gilbert}, {Dolphin}, {Fouesneau}, \& {Weisz}}]{Jennings2012}
{Jennings}, Z.~G., {Williams}, B.~F., {Murphy}, J.~W., {et~al.} 2012, \apj,
  761, 26

\bibitem[{{Jennings} {et~al.}(2014){Jennings}, {Williams}, {Murphy},
  {Dalcanton}, {Gilbert}, {Dolphin}, {Weisz}, \& {Fouesneau}}]{Jennings2014}
---. 2014, \apj, 795, 170

\bibitem[Karachentsev et al.(2000)]{Karachentsev2000} Karachentsev, I.~D., Sharina, M.~E., \& Huchtmeier, W.~K.\ 2000, \aap, 362, 544.

\bibitem[{{Kong} {et~al.}(2003){Kong}, {Sjouwerman}, {Williams}, {Garcia}, \&
  {Dickel}}]{Kong2003}
{Kong}, A.~K.~H., {Sjouwerman}, L.~O., {Williams}, B.~F., {Garcia}, M.~R., \&
  {Dickel}, J.~R. 2003, \apjl, 590, L21

\bibitem[Lacey \& Duric(2001)]{Lacey2001} Lacey, C.~K., \& Duric, N.\ 2001, \apj, 560, 719.

\bibitem[{{Lee} \& {Lee}(2014{\natexlab{a}})}]{Lee2014M31}
{Lee}, J.~H., \& {Lee}, M.~G. 2014{\natexlab{a}}, \apj, 786, 130

\bibitem[{{Lee} \& {Lee}(2014{\natexlab{b}})}]{Lee2014M33}
---. 2014{\natexlab{b}}, \apj, 793, 134

\bibitem[{{Long} {et~al.}(2014){Long}, {Kuntz}, {Blair}, {Godfrey},
  {Plucinsky}, {Soria}, {Stockdale}, \& {Winkler}}]{Long2014}
{Long}, K.~S., {Kuntz}, K.~D., {Blair}, W.~P., {et~al.} 2014, \apjs, 212, 21

\bibitem[{{Long} {et~al.}(2010{\natexlab{b}}){Long}, {Blair}, {Winkler},
  {Becker}, {Gaetz}, {Ghavamian}, {Helfand}, {Hughes}, {Kirshner}, {Kuntz},
  {McNeil}, {Pannuti}, {Plucinsky}, {Saul}, {T{\"u}llmann}, \&
  {Williams}}]{BlairLong2010}
---. 2010, \apjs, 187, 495

\bibitem[{{Maggi} {et~al.}(2016){Maggi}, {Haberl}, {Kavanagh}, {Sasaki},
  {Bozzetto}, {Filipovi{\'c}}, {Vasilopoulos}, {Pietsch}, {Points}, {Chu},
  {Dickel}, {Ehle}, {Williams}, \& {Greiner}}]{Maggi2016}
{Maggi}, P., {Haberl}, F., {Kavanagh}, P.~J., {et~al.} 2016, \aap, 585, A162

\bibitem[{{Marigo} {et~al.}(2008){Marigo}, {Girardi}, {Bressan}, {Groenewegen},
  {Silva}, \& {Granato}}]{marigo2008}
{Marigo}, P., {Girardi}, L., {Bressan}, A., {et~al.} 2008, \aap, 482, 883

\bibitem[{{Maund}(2017)}]{maund2017}
{Maund}, J.~R. 2017, \mnras, 469, 2202

\bibitem[{{Michael} {et~al.}(2003){Michael}, {McCray}, {Chevalier},
  {Filippenko}, {Lundqvist}, {Challis}, {Sugerman}, {Lawrence}, {Pun},
  {Garnavich}, {Kirshner}, {Crotts}, {Fransson}, {Li}, {Panagia}, {Phillips},
  {Schmidt}, {Sonneborn}, {Suntzeff}, {Wang}, \& {Wheeler}}]{Michael2003}
{Michael}, E., {McCray}, R., {Chevalier}, R., {et~al.} 2003, \apj, 593, 809

\bibitem[{{Millar} {et~al.}(2011){Millar}, {White}, {Filipovi{\'c}}, {Payne},
  {Crawford}, {Pannuti}, \& {Staggs}}]{Millar2011}
{Millar}, W.~C., {White}, G.~L., {Filipovi{\'c}}, M.~D., {et~al.} 2011, \apss,
  332, 221

\bibitem[{Pacholczyk(1970)}]{Pacholczyk1970}
Pacholczyk, A.~G. 1970, Series of Books in Astronomy and Astrophysics (San
  Francisco, CA: Freeman)

\bibitem[{{Pannuti}(2000)}]{Pannuti2000PhD}
{Pannuti}, T.~G. 2000, PhD thesis, University of New Mexico

\bibitem[{{Pannuti} {et~al.}(2000){Pannuti}, {Duric}, {Lacey}, {Goss},
  {Hoopes}, {Walterbos}, \& {Magnor}}]{Pannuti2000}
{Pannuti}, T.~G., {Duric}, N., {Lacey}, C.~K., {et~al.} 2000, \apj, 544, 780

\bibitem[{{Pannuti} {et~al.}(2017){Pannuti}, {Filipovi{\'c}}, {Luken}, {Wong},
  {Manojlovi{\'c}}, {Maxted}, \& {Roper}}]{Pannuti2017}
{Pannuti}, T.~G., {Filipovi{\'c}}, M.~D., {Luken}, K., {et~al.} 2017, Serbian
  Astronomical Journal, 195, 23

\bibitem[{{Pannuti} {et~al.}(2014){Pannuti}, {Rho}, {Heinke}, \&
  {Moffitt}}]{Pannuti2014}
{Pannuti}, T.~G., {Rho}, J., {Heinke}, C.~O., \& {Moffitt}, W.~P. 2014, \aj,
  147, 55

\bibitem[{{Payne} {et~al.}(2004){Payne}, {Filipovi{\'c}}, {Pannuti}, {Jones},
  {Duric}, {White}, \& {Carpano}}]{Payne2004}
{Payne}, J.~L., {Filipovi{\'c}}, M.~D., {Pannuti}, T.~G., {et~al.} 2004, \aap,
  425, 443

\bibitem[{{Read} {et~al.}(1997){Read}, {Ponman}, \& {Strickland}}]{Read1997}
{Read}, A.~M., {Ponman}, T.~J., \& {Strickland}, D.~K. 1997, \mnras, 286, 626

\bibitem[{{Reynolds}(2012)}]{Reynolds2012} Reynolds, S.~P., Gaensler, B.~M., \& Bocchino, F.\ 2012, \ssr, 166, 231.

\bibitem[{{Rodr{\'{\i}}guez} {et~al.}(2016){Rodr{\'{\i}}guez}, {Baume}, \&
  {Feinstein}}]{Rodriguez2016}
{Rodr{\'{\i}}guez}, M.~J., {Baume}, G., \& {Feinstein}, C. 2016, \aap, 594, A34

\bibitem[{{Seok} {et~al.}(2013){Seok}, {Koo}, \& {Onaka}}]{Seok2013}
{Seok}, J.~Y., {Koo}, B.-C., \& {Onaka}, T. 2013, \apj, 779, 134

\bibitem[{{Williams} {et~al.}(2014{\natexlab{a}}){Williams}, {Peterson},
  {Murphy}, {Gilbert}, {Dalcanton}, {Dolphin}, \& {Jennings}}]{williams2014a}
{Williams}, B.~F., {Peterson}, S., {Murphy}, J., {et~al.} 2014{\natexlab{a}},
  \apj, 791, 105

\bibitem[{{Williams} {et~al.}(2014{\natexlab{b}}){Williams}, {Lang},
  {Dalcanton}, {Dolphin}, {Weisz}, {Bell}, {Bianchi}, {Byler}, {Gilbert},
  {Girardi}, {Gordon}, {Gregersen}, {Johnson}, {Kalirai}, {Lauer}, {Monachesi},
  {Rosenfield}, {Seth}, \& {Skillman}}]{Williams2014}
{Williams}, B.~F., {Lang}, D., {Dalcanton}, J.~J., {et~al.} 2014{\natexlab{b}},
  \apjs, 215, 9

\bibitem[{{Winkler} {et~al.}(2017){Winkler}, {Blair}, \& {Long}}]{Winkler2017}
{Winkler}, P.~F., {Blair}, W.~P., \& {Long}, K.~S. 2017, \apj, 839, 83

\end{thebibliography}





\label{lastpage}
\end{document}